\documentclass[%
 aip,jcp,superscriptaddress,
 amsmath,amssymb,floatfix,longbibliography,
 reprint,%
]{revtex4-2}

\usepackage{graphicx}
\usepackage{dcolumn}
\usepackage{bm}
\usepackage[version=4]{mhchem}

\usepackage[utf8]{inputenc}
\usepackage[T1]{fontenc}
\usepackage{mathptmx}
\usepackage{etoolbox}
\usepackage{braket}
\usepackage{xcolor}
\usepackage{url}
\usepackage[hidelinks]{hyperref}

\usepackage[normalem]{ulem}

\makeatletter
\def\@email#1#2{%
 \endgroup
 \patchcmd{\titleblock@produce}
  {\frontmatter@RRAPformat}
  {\frontmatter@RRAPformat{\produce@RRAP{*#1\href{mailto:#2}{#2}}}\frontmatter@RRAPformat}
  {}{}
}%
\makeatother
\begin{document}

\preprint{AIP/123-QED}

\title[]{Ultrafast laser-driven quantum dynamics in positronium chloride}
\author{Einar Aurbakken}
\email{einar.aurbakken@kjemi.uio.no}
\affiliation{Hylleraas Centre for Quantum Molecular Sciences, Department of Chemistry, University of Oslo, P.O. Box 1033 Blindern, N-0315 Oslo, Norway}%
\author{H{\aa}kon Emil Kristiansen}
\affiliation{Hylleraas Centre for Quantum Molecular Sciences, Department of Chemistry, University of Oslo, P.O. Box 1033 Blindern, N-0315 Oslo, Norway}%
\author{Simen Kvaal}
\affiliation{Hylleraas Centre for Quantum Molecular Sciences, Department of Chemistry, University of Oslo, P.O. Box 1033 Blindern, N-0315 Oslo, Norway}%
\author{Antoine Camper}
\affiliation{Department of Physics, University of Oslo, P.O. Box 1048 Blindern, N-0316 Oslo, Norway}
\author{Thomas Bondo Pedersen}
\email{t.b.pedersen@kjemi.uio.no}
\affiliation{Hylleraas Centre for Quantum Molecular Sciences, Department of Chemistry, University of Oslo, P.O. Box 1033 Blindern, N-0315 Oslo, Norway}%

\date{\today}

\begin{abstract}
We present a computational study of the laser-driven quantum dynamics of positronium (Ps), PsH, and PsCl at the time-dependent Hartree-Fock level of theory. To eliminate finite-basis effects and to properly capture continuum dynamics, we use a spherical polar pseudospectral representation.
The multicomponent theory and its implementation are described in detail.
We find that while the presence of the positron delays electron ionization in PsH, a slight enhancement of electron ionization is observed in PsCl.
In both cases, the positronic response is faster than that of the electrons.
We propose that the formation of PsCl may be directly observed through photopositron spectra in the multiphoton regime, where PsCl peaks are expected at roughly twice the energy of Ps peaks, making PsCl clearly distuinguishable from Ps.
In the tunelling regime, however, photopositron rescattering peaks may only be distuinguishable if the amount of Ps is sufficiently low.
\end{abstract}

\maketitle

\section{Introduction}

When an electron comes into contact with its antiparticle, the positron, the pair decays into 
(typically) two gamma photons
at $511\,\text{keV}$ each.
This quantum electrodynamical positron annihilation process is crucial for medical diagnostics
through positron emission tomography and its recent extension to
positronium imaging (PI).~\cite{moskal_positronium_2019,bass_colloquium_2023,moskal_positronium_2025}
The PI technique exploits the formation of short-lived positronium (Ps) ``atoms'',
the hydrogen-like state of a positron and an electron bound by their Coulomb interaction, inside the body.
The key feature exploited for imaging is the strong dependence of the lifetime on the surroundings of Ps.
Similarly, positron annihilation lifetime spectroscopy has evolved into an important characterization technique
in chemistry and material science.~\cite{bass_colloquium_2023,jean_positron_1990,dlubek_positron_2002,gidley_positron_2006,singh_positron_2016,attallah_revisiting_2024}

Despite the advanced medical and scientific applications of positron annihilation, the bound states formed
by positrons and atoms or molecules have proven very challenging to study theoretically.
This poses a problem for further development of positron-based technologies and for fundamental
scientific investigations of, e.g., exotic positron-mediated chemical bonding.~\cite{Charry18,moncada_covalent_2019,Charry22,Archila24}
The main obstacle is the high-level many-body theory required to accurately describe electronic and electron-positron
correlations and to predict positron affinities of atoms and molecules.~\cite{Hofierka22,charry_martinez_correlated_2022,upadhyay_capturing_2024,tumakov_relativistic_2024}
Experimentally, molecular positron affinities can be measured indirectly through vibrational Feshbach resonances~\cite{Gribakin02,gribakin_positron-molecule_2010} while alternative techniques have been proposed for atoms.~\cite{mitroy_measuring_1999,Surko_2012}
However, with lifetimes on the order of nanoseconds,
it should be possible to study positron--matter binding and dynamics directly by ultrafast spectroscopy.

The development of femtosecond and attosecond laser pulses and associated technologies have paved the way for new research fields in chemistry and physics.~\cite{Corkum2007,Krausz2009,Nisoli2017} Different effects including impulsive stimulated  scattering~\cite{Yan1985} have been discovered and new spectroscopic methods such as time-resolved pump-probe techniques have been developed. In addition, generation of strong-field pulses have revealed a plethora of ultrafast strong-field phenomena such as non-sequential double 
ionization,~\cite{Fittinghoff92,Walker94} recollision-driven inner-shell processes,~\cite{Deng16} emergence of plateaus in 
above-threshold ionization,~\cite{Becker2018} laser-induced electron diffraction,~\cite{Blaga12} and high-order harmonic generation.~\cite{McPherson87,Ferray88}
These phenomena are generally well-described in the strong-field approximation~\cite{Lewenstein94} and have given birth to the field of attochemistry.~\cite{Merritt21} Laser-driven recollision processes have also been proposed for driving fusion~\cite{Ditmire99} in clusters and for muon pair creation in Ps.~\cite{Muller08}

In principle, laser interaction with positron binding atoms and molecules should also entail a large variety of ultrafast phenomena resembling the ones mentioned above, albeit with major differences due to the presence of the bound positron. The positron, located further out, will have a shielding effect on the electrons, altering the electron dynamics relative to the atomic or ionic counterpart system. Also, the positron itself will entail its own dynamics. One possibility for the positron is to tunnel ionize through the Coulombic potential and play the role of the rescattering particle. However, the numerical study of these phenomena is so far limited by the existing computational tools. 

In 2018, \citet{Suzuki18} reported an implementation of multicomponent time-dependent density-functional theory for coupled electron-positron dynamics driven by a laser field. Using a plane-wave basis with norm-conserving pseudopotentials, they presented an application to $e^+$--\ce{LiH} exposed to weak gaussian-shaped laser pulses with a peak intensity of
$0.53 \times 10^{12}\,\mathrm{W/cm^2}$ (electric-field strength $0.2\,\mathrm{V/}${\AA})
and wavelengths of $413$, $827$, and $2480\,\mathrm{nm}$ (frequencies $0.5$, $1.5$, and $3.0\,\mathrm{eV}$),
showing that the presence of the positron suppresses a bound electronic excitation 
in \ce{LiH}. Moreover, they found that the quivering motion of the electron density in $e^+$--\ce{LiH} is opposite to that in
\ce{LiH}. These effects were ascribed to dynamical electron-positron correlation.

Here, we neglect dynamical correlation and use all-particle grid-based time-dependent Hartree-Fock (TDHF) theory to shed light on the time evolution of positronium chloride (PsCl) exposed to external laser fields. 
Although the target quantity of interest in most ground-state studies, namely the positron binding energy, requires a high-level treatment of correlation effects (including large basis-set expansions) to achieve sufficient accuracy, the recent work by
\citet{Leveque24} indicates that the PsCl orbitals are qualitatively correct at the Hartree-Fock level.
Extending this observation, we thus implicitly assume that the laser-induced quantum dynamics can be captured qualitatively correctly at the TDHF level of theory.
Importantly, we effectively eliminate the basis-set issue,~\cite{Bromley2001, Bromley2006} both in the ground state and for the dynamical evolution of the wavefunction, by using a pseudospectral radial grid and angular expansions in spherical harmonics. Thus, high angular momenta are still needed for a proper description of the time-dependent wave function of systems subject to strong laser fields as well as for describing the ground state if correlated methods are employed, see, for example, the works by~\citeauthor{saito2003_psh_mrci}.~\cite{saito2003_psh_mrci, saito2003_mrci_ps_halides}
The present work lays the foundations for correlated treatments using grid-based methodologies.

In this paper, we give a detailed account of TDHF theory for two-component systems in a spherical polar pseudo-spectral representation, including an exhaustive description of the integrator used to propagate the nonlinear TDHF equations. Rather than the density matrix, we focus on orbital time evolution to facilitate future extensions to correlated methods. 
Qualitative aspects of the laser-induced coupled electron-positron dynamics are discussed and related to the work of
\citeauthor{Suzuki18}.~\cite{Suzuki18}
We compute above-threshold ionization (ATI) spectra for the positron in both the multiphoton regime and in the tunneling regime, thus indicating a potential ultrafast spectroscopic method to detect PsCl.
Finally, we compare TDHF results with those obtained from a ``single active positron'' (SAP) model where the electronic density is kept frozen at its ground-state distribution during the laser-driven dynamics. The SAP model thus is similar to
the single active electron (SAE)\cite{Kulander1988,Schafer90,Kulander1993} models that are widely used to study ultrafast laser-induced processes in electronic systems.

\section{Theory}
\subsection{Model and computational method}
\label{sec:model_and_computational_method}
Unless stated otherwise, we use atomic units (a.u.) throughout this paper.
The clamped-nuclei Born-Oppenheimer Hamiltonian~\cite{Born1927,Born1954} in a.u. of $N_e$ electrons (mass $1\,\mathrm{a.u.}$, charge $-1\,\mathrm{a.u.}$) and a single positron (mass $1\,\mathrm{a.u.}$, charge $1\,\mathrm{a.u.}$) interacting with a time-dependent uniform electric field is given by
\begin{align}
    \hat{H} &= \hat{h}^p(\mathbf{r}_p, t) + \sum_{i=1}^{N_e} \hat{h}^e(\mathbf{r}_i, t)  \nonumber \\
    &+ \sum_{i < j} \frac{1}{|\mathbf{r}_i-\mathbf{r}_j|}  - \sum_{i=1}^{N_e} \frac{1}{|\mathbf{r}_p-\mathbf{r}_i|}, \label{electron_positron_Hamiltonian}
\end{align}
with $\mathbf{r}_i$ ($\mathbf{r}_j$) and $\mathbf{r}_p$ being the spatial coordinates of the $i$th ($j$th) electron and the positron, respectively. 
The term $1/|\mathbf{r}_i-\mathbf{r}_j|$ is the \textit{repulsive} Coulomb interaction between the $i$th and $j$th electrons, while $-1/|\mathbf{r}_p-\mathbf{r}_i|$ is the \textit{attractive} Coulomb interaction between the positron and the $i$th electron.

Within the velocity-gauge electric-dipole approximation, the single-electron and single-positron contributions to the Hamiltonian are given by
\begin{align}
    \hat{h}^e(\mathbf{r}, t) &= -\frac{1}{2} \nabla^2 -  V_{\text{ext}}(\mathbf{r}) - \mathrm{i}A(t) \mathbf{u} \cdot \nabla, \\
    \hat{h}^p(\mathbf{r_p}, t) &= -\frac{1}{2} \nabla_p^2 +  V_{\text{ext}}(\mathbf{r}_p) + \mathrm{i}A(t) \mathbf{u} \cdot \nabla_p,
\end{align}
where
\begin{equation}
    V_{\text{ext}}(\mathbf{r}) = \sum_{\alpha=1}^M \frac{Z_\alpha}{|\mathbf{r}-\mathbf{R}_{\alpha}|},
\end{equation}
is the potential due to $M$ nuclei situated at $\{\mathbf{R}_\alpha\}_{\alpha=1}^M$ with charge $\{ Z_\alpha \}_{\alpha=1}^M$.
For PsCl we have $N_e=18$, $M=1$, $Z_\mathrm{Cl}=17$, and $\mathbf{R}_\mathrm{Cl}=\mathbf{0}$, while for PsH we have
$N_e=2$, $M=1$, $Z_\mathrm{H}=1$, and $\mathbf{R}_\mathrm{H}=\mathbf{0}$.

The linearly polarized vector potential $A(t)\mathbf{u}$, where $\mathbf{u}$ is a real unit polarization vector,
is related to the electric-field amplitude $\mathcal{E}(t)$ as
\begin{equation}
    \label{eq:vector_potential}
    A(t) = -\int_0^t \mathcal{E}(t') \mathrm{d}t',
\end{equation}
where we have assumed $\mathcal{E}(0) = 0$.

In the Hartree-Fock approximation (with the presence of a \textit{single} positron), the wavefunction is the tensor product of a single $N_e$-electron Slater determinant and a single positron orbital~\cite{cade1977electronic, kurtz1980theoretical}
\begin{equation}
    \Psi(\mathbf{r}_1, \cdots, \mathbf{r}_{N_e}, \mathbf{r}_p, t) = \Phi(\mathbf{r}_1, \cdots, \mathbf{r}_{N_e}, t) \otimes \psi(\mathbf{r}_p, t).
\end{equation}
We assume that the electron orbitals are doubly occupied---i.e., we use closed-shell spin-restricted Hartree-Fock (RHF) theory---such that the number of spatial orbitals is $N_e/2$.

The orbital equations of motion are derived using the stationary-action time-dependent variational principle.
With the canonical gauge choice,
the equations of motion for the time-dependent electron orbitals $\{ \phi_i(\mathbf{r}, t) \}_{i=1}^{N_e/2}$ and the positron orbital $\psi(\mathbf{r}_p, t)$ become the (canonical) TDHF equations
\begin{align}
    &\mathrm{i}\partial_t \phi_i(\mathbf{r}, t) = \hat{f}^{e}(\{\phi_i\}, \psi, t) \phi_i(\mathbf{r}, t), \\
    &\mathrm{i}\partial_t \psi(\mathbf{r}_p,t) = \hat{f}^{p}(\{\phi_i\}, \psi, t) \psi(\mathbf{r}_p, t).
\end{align}
The electron and positron Fock operators $\hat{f}^e$ and $\hat{f}^p$ are given by
\begin{align}
    &\hat{f}^{e}(\{\phi_i\}, \psi,t) \phi_i(\mathbf{r},t) = \hat{h}^e(\mathbf{r},t)\phi_i(\mathbf{r}, t)  \nonumber \\ 
    &+ \left(2 V^{\text{dir}, e}(\mathbf{r}, t) - V^{\text{dir}, p}(\mathbf{r}, t) \right)\phi_i(\mathbf{r}, t) \nonumber \\
    &- \sum_{j=1}^{N_e/2} V^{\text{exc}, e}_{j,i}(\mathbf{r},t) \phi_j(\mathbf{r},t) , \\
    &\hat{f}^{p}(\{\phi_i\}, \psi, t) \psi(\mathbf{r}_p, t) = \left(\hat{h}^p(\mathbf{r}_p,t) - 2 V^{\text{dir}, e}(\mathbf{r}, t) \right) \psi(\mathbf{r}_p, t),
\end{align}
where we have introduced the direct and exchange potentials
\begin{align}
    V^{\text{dir}, e}(\mathbf{r}, t) &= \sum_{j=1}^{N_e/2} \int \frac{|\phi_j(\mathbf{r}', t)|^2}{|\mathbf{r}-\mathbf{r}'|} \mathrm{d}\mathbf{r}', \\
    V^{\text{dir}, p}(\mathbf{r}, t) &= \int \frac{|\psi(\mathbf{r}_p, t)|^2}{|\mathbf{r}-\mathbf{r}_p|} \mathrm{d}\mathbf{r}_p, \\
    V^{\text{exc}, e}_{j,i}(\mathbf{r},t) &= \int \frac{\phi_j^*(\mathbf{r}', t) \phi_i(\mathbf{r}', t)}{|\mathbf{r}-\mathbf{r}'|} \mathrm{d}\mathbf{r}'. 
\end{align}

The initial state is taken as the RHF ground state determinant of the field-free Hamiltonian ($\mathcal{E}(t)=0$), which is found by solving the stationary canonical RHF equations
\begin{align}
    &\hat{f}^e(\{\phi_i\}, \psi, t=0) \phi_i(\mathbf{r}) = \epsilon_i^e \phi_i(\mathbf{r}), \\
    &\hat{f}^{p}(\{\phi_i\}, \psi, t=0) \psi(\mathbf{r}_p) = \epsilon_0^p \psi(\mathbf{r}_p),
\end{align}
for the occupied electron and positron orbitals.

In the absence of an external electric field, the Hartree-Fock potential (direct plus exchange) for atoms or ions with closed subshells (such as \ce{Cl-}) is spherically symmetric.~\cite{bransden2003physics} This also holds for the direct potential due to the positron. Hence, to solve the HF equations for the initial state, we use spherical coordinates and assume that the orbitals are angular-momentum eigenfunctions of the form
\begin{align}
    \phi_i(r,\theta,\varphi) &= r^{-1}u^i_{n,l}(r)Y_{l,m}(\theta, \varphi), \\
    \psi(r,\theta,\varphi) &= r^{-1}v_{n,l}(r)Y_{l,m}(\theta, \varphi).
\end{align}

For a linearly polarized electric field, the total potential becomes cylindrically symmetric
and taking the polarization direction parallel to the $z$-axis, the interaction of the electrons or positron with the electric field can, in spherical coordinates, be written as
\begin{equation}
    V(r,\theta, \varphi, t) = \mathrm{i} qA(t) \left[\cos{\theta}\frac{\partial}{\partial r} - \frac{\sin{\theta}}{r}\frac{\partial}{\partial \theta}\right].
\end{equation}
With this polarization, the initial quantum numbers $m$ of the electron and positron orbitals are conserved,~\cite{sato2016time} and a time-dependent orbital can be expanded as 
\begin{align}
    &\phi_i(r,\theta,\varphi,t) = \sum_{l=0}^{l_{\text{max}}} r^{-1} u^{i}_{l,m_i}(r,t) Y_{l,m_i}(\theta,\varphi), \\
    \label{eq:positron_expansion}
    &\psi(r,\theta,\varphi,t) = \sum_{l=0}^{l_{\text{max}}} r^{-1} v_{l,0}(r,t) Y_{l,0}(\theta,\varphi),
\end{align}
where $l_{\text{max}}$ is a finite cut-off value for the number of angular momenta included in the expansion. 
As each orbital has fixed $m$, we will in the following use the short-hand notation $u_l^i(r,t) = u_{l,m_i}^i(r,t)$ and $v_l(r,t) = v_{l,0}(r,t)$. We note that the radial functions satisfy homogeneous Dirichlet boundary conditions, i.e., $u^i_{l,m}(r,t)$ and $v_{l,0}(r,t)$ vanish at $r=0$ and as $r \rightarrow \infty$ for all $t$.

For the Coulomb interaction operator, we employ the multipole expansion~\cite{bransden2003physics}
\begin{align}\label{eq:r12_expansion}
    \frac{1}{\vert \mathbf{r}_1 - \mathbf{r}_2 \vert} \approx
    \sum_{L=0}^{L_{\text{max}}} \sum_{M=-L}^{L} \frac{4\pi}{2L+1}\frac{r_<^L}{r_>^{L+1}}Y_{L,M}^*(\Omega_1)Y_{L,M}(\Omega_2),
\end{align}
where $\Omega_i = (\theta_i, \varphi_i)$, $i=1,2$,
\begin{equation}
    r_< = \min(r_1, r_2), \qquad
    r_> = \max(r_1, r_2),
\end{equation}
and $L_{\text{max}}$ is a finite cut-off value. 

Insertion of the above expansions into the TDHF equations and integrating out angular variables results in equations of motion for the radial part of the electron orbitals 
\begin{align}
    &\mathrm{i} \partial_t u_{l}^i(r,t) =  -\frac{1}{2} \Delta_l  u_{l}^i(r,t) - \frac{Z}{r}u_{l}^i(r,t) \nonumber \\
    &-\mathrm{i}A(t) [ a_{l-1, m_i}D_{-l}u_{l-1}^i(r,t) + a_{l, m_i}D_{l+1} u_{l+1}^i(r,t) ]\nonumber \\
    &+2\sum_{l'} \mathfrak{D}_{l,l'}^{m_i}(r) u_{l'}^i(r,t) -\sum_{l'} \mathfrak{P}_{l,l'}^{m_i}(r) u_{l'}^i(r,t) \nonumber \\
    &- \sum_{j} \sum_{l'} \mathfrak{X}_{ij, ll'}^{m_im_j}(r) u_{l'}^j(r,t) \label{eq:radial_eom_electron_orbs},
\end{align}
and the positron orbital
\begin{align}
    &\mathrm{i} \partial_t v_{l}(r,t) =  -\frac{1}{2} \Delta_l  v_{l}(r,t) + \frac{Z}{r}v_{l}(r,t) \nonumber \\
    &+\mathrm{i}A(t) [ a_{l-1, 0}D_{-l}v_{l-1}(r,t) + a_{l, 0}D_{l+1} v_{l+1}(r,t) ]\nonumber \\
    &-2\sum_{l'} \mathfrak{D}_{l,l'}^{0}(r) v_{l'}(r,t) \label{eq:radial_eom_positron_orbs}.
\end{align}
Here, we have used the following definitions
\begin{align}
    &a_{l,m} = \sqrt{\frac{(l+1)^2-m^2}{(2l+1)(2l+3)}}, \\ 
    &D_l = \frac{\partial}{\partial r} + \frac{l}{r}, \\
    &\Delta_l = \frac{\partial^2}{\partial r^2} - \frac{l(l+1)}{r^2}.
\end{align}
The radial forms of the direct and exchange matrices are
\begin{align}
    &\mathfrak{D}_{l,l'}^{m_i}(r) = \sum_j \sum_{L} \sum_{l_1l_2} \mathfrak{u}_{l_1l_2,L}^{j,j} (r) g_{l_1Ll_2}^{m_jm_j}g_{lLl'}^{m_im_i}, \\
    &\mathfrak{P}_{l,l'}^{m_i}(r) =\sum_{L} \sum_{l_1l_2} \mathfrak{v}_{l_1l_2,L} (r) g_{l_1Ll_2}^{00}g_{lLl'}^{m_im_i}, \\
    &\mathfrak{X}_{ij, ll'}^{m_im_j}(r) = \sum_L \sum_{l_1l_2} \mathfrak{u}_{l_1l_2,L}^{j,i} (r)g_{l_1Ll_2}^{m_jm_i}g_{lLl'}^{m_im_j}(-1)^{m_j-m_i},
\end{align}
where we have defined the radial components of the direct and exchange potentials as
\begin{align}
    \mathfrak{u}_{l_1l_2,L}^{j,i}(r) = \frac{4\pi}{2L+1} \int  \frac{r_<^L}{r_>^{L+1}} u_{l_1}^{j*}(r')u_{l_2}^i (r')\mathrm{d}r', \label{eq:radial_dir_exc_pot_electrons}\\
    \mathfrak{v}_{l_1l_2,L}(r) = \frac{4\pi}{2L+1} \int  \frac{r_<^L}{r_>^{L+1}} v_{l_1}^{*}(r')v_{l_2} (r')\mathrm{d}r' \label{eq:radial_dir_pot_positron},
\end{align}
and an array of Gaunt-like coefficients
\begin{equation}
    g_{l_1 L l_2}^{m_i m_j} = \int Y_{l_1,m_i}^*(\Omega)Y_{L,(m_i-m_j)}(\Omega)Y_{l_2,tm_j}(\Omega) \mathrm{d}\Omega.
\end{equation}

The radial coordinate is discretized using the Gauss-Legendre-Lobatto (GLL) pseudospectral method (see Appendix~\ref{app:pseudospectral_grid}), and the ground-state RHF equations are solved numerically following the procedure described in Ref.~\citenum{cinal2020highly}. 

The radial direct and exchange potentials (Eqs.~\eqref{eq:radial_dir_exc_pot_electrons} and \eqref{eq:radial_dir_pot_positron}) satisfy a Poisson equation, which is solved on the GLL grid
following the procedure in Ref.~\citenum{hochstuhl2014time}. In principle, one could have computed the direct and exchange potentials by quadrature. However, due to presence of the derivative singularity at $r=r^\prime$, superior numerical accuracy is achieved by solving the Poisson equation.~\cite{hochstuhl2014time}

For the time propagation, we use the approximate second-order constant density matrix (CDM2) integration scheme
described in detail in Appendix~\ref{app:propagator}. 

The electric-field amplitude, representing a laser pulse with carrier frequency $\omega$ and zero carrier-envelope phase, is modeled by a sine-wave modulated by a trigonometric envelope function~\cite{Barth2009}
\begin{equation}\label{eq:electric_field}
    \mathcal{E}(t) = \mathcal{E}_0 \sin^2{\left(\pi \frac{t}{t_d}\right)}\sin(\omega t), \quad 0 \leq t \leq t_d,
\end{equation}
where $\mathcal{E}_0$ is the maximum field strength such that the peak intensity is given by $(1/2)\epsilon_0 c \mathcal{E}_0^2$, and $t_d$ is the (foot-to-foot) duration of the laser pulse.
All simulations are done in velocity gauge with the vector potential obtained from the electric field according to Eq.~\eqref{eq:vector_potential}.
The electric field is turned on at $t=0$ and turned off after $t_d = n_{\text{c}} t_\text{c}$, where $t_\text{c} = 2\pi/\omega$ is one \emph{optical cycle} and $n_{\text{c}}$ the number of optical cycles.

Finally, we will also do simulations of Ps. In this case we separate the two-particle Hamiltonian into a center-of-mass term, which does not couple to the electric field in the electric-dipole approximation, and an internal single-particle (hydrogen-like) Hamiltonian with the coordinate origin at the electron (i.e., we use the relative coordinate $\bm{r} = \bm{r}_p - \bm{r}_e$) and with reduced mass $\mu = 1/2$ and charge $q=1$. The internal Hamiltonian for Ps in an external electric field thus is given by
\begin{equation}
    \hat{H}_{\mathrm{Ps}} = \left( \bm{p} - A(t)\bm{u} \right)^2 - \frac{1}{r}.
\end{equation}
The time-dependent Schr{\"o}dinger equation for Ps is solved numerically using the same pseudo-spectral discretization as for the TDHF equations, with the wavefunction expansion in Eq.~\eqref{eq:positron_expansion} and with the ground-state wavefunction as initial condition.

\subsection{Above-threshold ionization spectra}

To investigate positron ionization we compute ATI spectra
using the procedure proposed by \citeauthor{Schafer90}.~\cite{Schafer90} That is, we calculate expectation values of the operator
\begin{equation}
    \hat{W}(E; \Gamma) = \frac{\Gamma^4}{(\hat{H} - E)^4 + \Gamma^4},
\end{equation}
in the final positronic state $\ket{\psi_f}$ after interaction with the laser pulse
such that the positron yield for an energy $E$ is given by $\bra{\psi_f}\hat{W}(E;\Gamma)\ket{\psi_f}$.
Here, $\Gamma$ is a broadening parameter and $\hat{H}$ is the Hamiltonian whose expectation value gives the energy of the positron as $t\rightarrow \infty$.
For PsCl, $\hat{H}$ is the Fock operator for the positron, while for Ps we assume equal kinetic energies of the electron and the positron after ionization
and thus $\hat{H} = \frac{1}{2} \hat{H}_{\mathrm{Ps}}$.

Since the yield for each energy in the generated spectrum is essentially a weighted sum of contributions from eigenstates of the Hamiltonian close in energy, the spectrum depends heavily on how well the eigenstates match the actual continuum states. If $r_{\text{max}}$ is chosen too small, the ATI spectrum will appear noisy even though the simulation parameters are well converged. We deal with this in two ways. The first way is to increase the broadening parameter $\Gamma$, as the averaging of contributions from nearby states will ultimately remove the noise. However, this also reduces the resolution of the spectrum. The second way is to map the final wavefunction onto a larger grid prior to the energy analysis. We do this using barycentric polynomial interpolation.

\section{Results}

\subsection{Computational details}

For the radial grids, we use $N_r = 1.25 r_\mathrm{max}$ grid points in all simulations, as we have found this to be sufficient to get converged results. 
We minimize reflections by converging the simulations with respect to $r_\mathrm{max}$ and, in addition, we use a mask function of the form~\cite{sato2016time}
\begin{align}
    M(r) &=
    \begin{cases}
        1, & r < r_0 \\
        \cos^{1/4} \left( \dfrac{\pi}{2} \dfrac{r - r_0}{r_{\max} - r_0} \right), & r_0 \leq r < r_{\max}
    \end{cases},
\end{align}
where we choose $r_0 = r_\mathrm{max} - 30\,\mathrm{a}_0$.
For the spherical expansions, we use odd values of $l_{\text{max}}$, and $L_{\text{max}} = (l_{\text{max}}-1)/2$.
Time steps are $\Delta t=0.1\,\text{a.u.}$ unless otherwise stated.

\subsection{Initial state}

\begin{figure}[h]
\includegraphics[width=8.6cm]{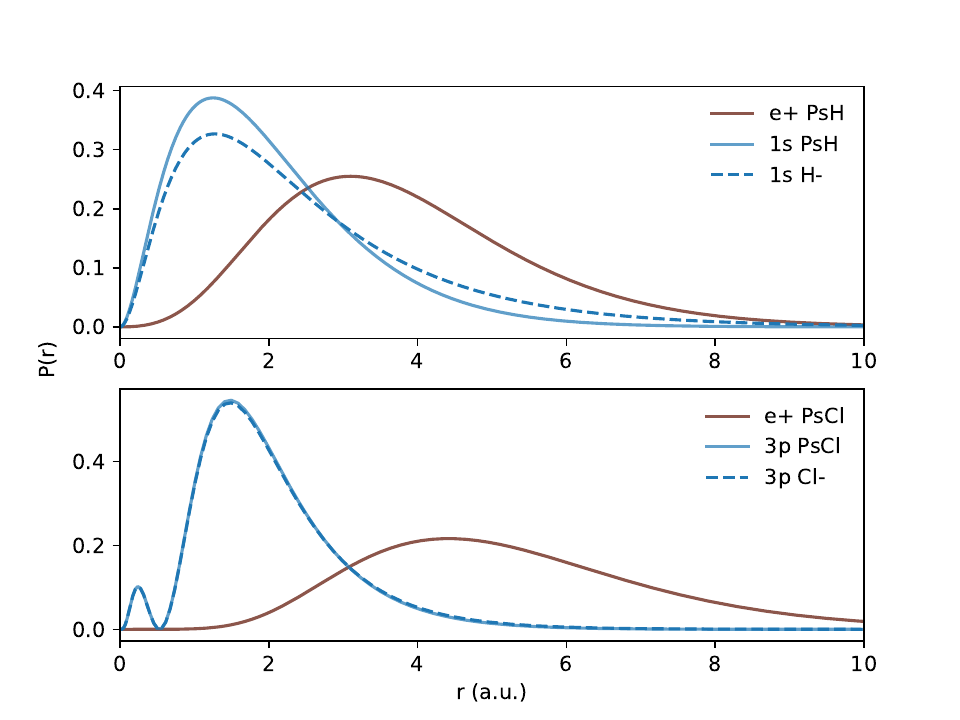}
\caption{\label{fig:gs_orbitals_A-vsPsA} Radial distribution functions for the positron orbital and the outermost electron orbitals of PsH (top) and PsCl (bottom).}
\end{figure}

The ground state of PsCl is an S state (total angular momentum $0$).
In agreement with the results reported by \citet{Saito2000} for PsH using Hylleraas functions, the addition of a spherically symmetric attractive Coulomb potential from the positron squeezes the electrons toward the atomic nucleus.
This effect, however, is much less pronounced for PsCl than for PsH, as can be seen in Fig.~\ref{fig:gs_orbitals_A-vsPsA}.
The lower-level sub-shells are left out of the figure since the orbitals from PsCl and Cl$^-$ are indistinguishable at this scale.

To quantify the effect of the positron on the electronic orbitals, we define the orbital relaxation energy
\begin{align}
    \Delta E^\mathrm{relax}_\mathrm{PsA} = E_\mathrm{Ps-A} - E^\mathrm{HF}_\mathrm{PsA},
\end{align}
along with the relative orbital relaxation energy,
\begin{align}
    \delta E^\mathrm{relax}_\mathrm{PsA} = \frac{\Delta E^\mathrm{relax}_\mathrm{PsA}}{E^\mathrm{HF}_\mathrm{PsA}},
\end{align}
where $E^\mathrm{HF}_\mathrm{PsA}$ is the full self-consistent HF energy of PsA , while $E_\mathrm{Ps-A}$ is obtained by diagonalizing $\hat{f}^p$ in the mean-field of A$^-$ and creating a product state of the A$^-$ ground-state Slater determinant and the resulting positron orbital. 

We find $\Delta E^\mathrm{relax}_\mathrm{PsH} = 14\,\mathrm{mE}_h$ and $\Delta E^\mathrm{relax}_\mathrm{PsCl} = 1.2\,\mathrm{mE}_h$,
 while the relative values are $\delta E^\mathrm{relax}_\mathrm{PsH} = 2.1\%$ and $\delta E^\mathrm{relax}_\mathrm{PsCl} = 2.5 \times 10^{-4}\%$. This confirms that the effect of the positron is much smaller on the electrons in PsCl than in PsH, not just in relative terms but also in absolute terms. At least in the mean-field model, the addition of the positron to Cl$^-$ constitutes a small perturbation of the electronic system.

\subsection{Qualitative assessment of the dynamics}

Although the addition of a positron to Cl$^-$ has little effect on the ground-state electronic orbitals, the situation is markedly different for laser-induced dynamics. Because of the addition of a light positive charge, the dynamics in PsA systems are more complex than in ordinary atoms. The electrons are bound to a central attractive potential, whereas the positron is bound only due to the presence of the electrons. This breaks the symmetry between the electron and the positron present in isolated Ps. 

\begin{figure}[h]
\includegraphics[width=8.6cm]{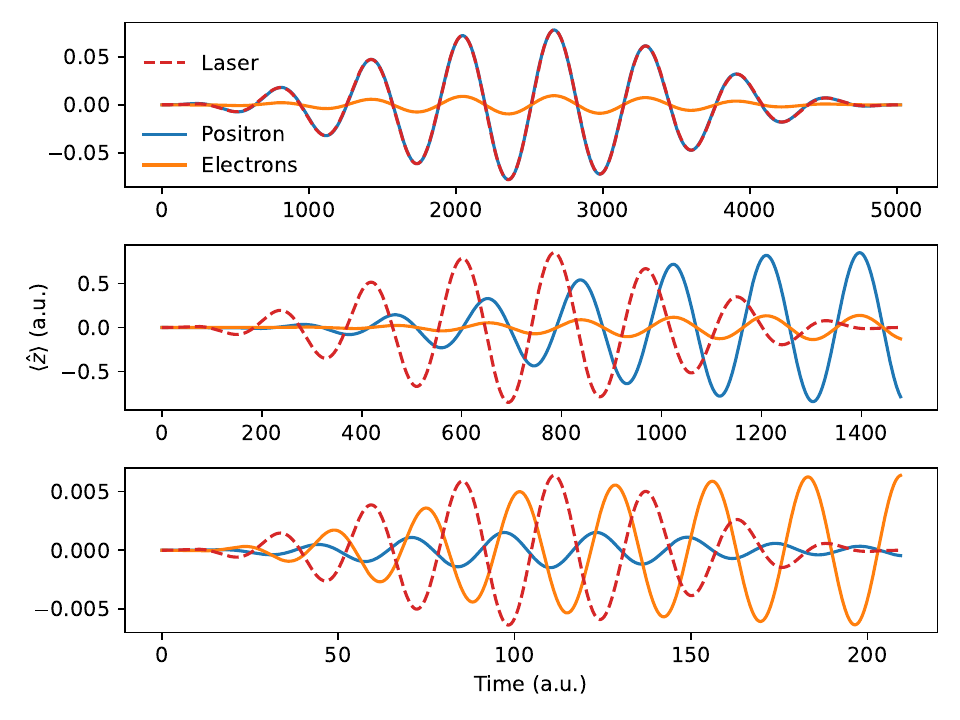}
\caption{\label{fig:dipole_laser_phase} Position expectation value for electrons and positron in PsCl subject to a weak ($\mathcal{E}_0=0.0001\,\mathrm{a.u.}$) 8-cycle laser pulse with frequencies $0.01\,\mathrm{a.u.}$ (top), $0.034\,\mathrm{a.u.}$ (middle) and $0.24\,\mathrm{a.u.}$ (bottom). These correspond, in order, to a frequency in the transparent region, the first positron resonance, and the first electronic resonance. The units of the laser pulse are arbitrary, as its meant to show the phase between the electric field and the position expectation values. The grid parameters were $r_{\text{max}}=100$ and $l_{\text{max}}=9$.}
\end{figure}

\begin{figure}[h]
\includegraphics[width=8.6cm]{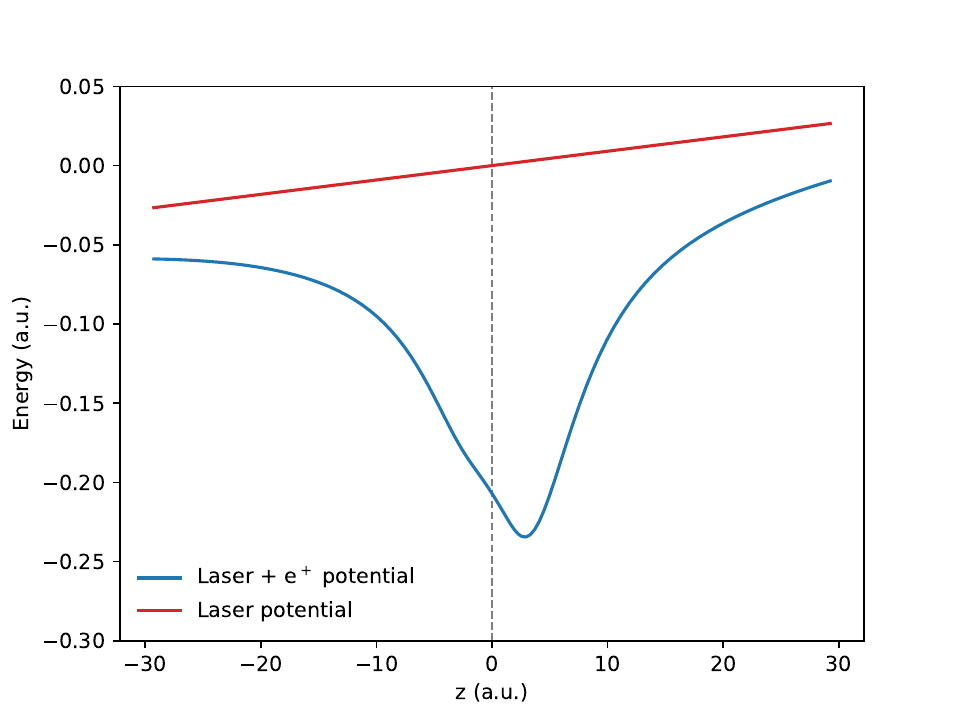}
\caption{\label{fig:external_potential_seen_by_e-_0576} The external potential (excluding nucleus) seen by the electrons in a Cl$^-$ system with and without a positron.}
\end{figure}

According to Ehrenfest's theorem~\cite{Ehrenfest1927} 
\begin{equation}
    \label{eq:ehrenfest}
    \frac{\mathrm{d}^2}{\mathrm{d} t^2} \braket{\hat{z}} = q\mathcal{E}(t) - \braket{\frac{\partial V}{\partial \hat{z}}},
\end{equation}
where $\hat{z}$ is the position operator of either the electrons ($q=-1$) or the positron ($q=+1$) and $V$ denotes the total Coulomb operator,
the laser-induced dynamics become a balance of forces stemming from the external laser field and the attractive mean-field Coulomb forces among the electrons and the positron.
\citet{Suzuki18} noted that in the e$^+$-LiH system, the electron moves in the same direction as the external field, opposite to what is expected from the first
term in Eq.~\eqref{eq:ehrenfest}. For weak, low-frequency pulses, we observe exactly the same in PsCl, as illustrated by the top panel in Fig.~\ref{fig:dipole_laser_phase}.
The fundamental reason for this is that the positron is more loosely bound than the electrons and consequently exhibits a stronger response to the external field. The attractive electrostatic field set up by the displaced positron locally dominates the potential due to the laser field, essentially making the electrons follow the motion of the positron. Fig.~\ref{fig:external_potential_seen_by_e-_0576} shows a snapshot of this effect at $t=2600\,\mathrm{a.u.}$ for the run corresponding to the top panel in Fig.~\ref{fig:dipole_laser_phase}, but with field strength increased to $\mathcal{E}_0=0.001\,\mathrm{a.u.}$ to make the effect more easily visible on the figure.
 
While the external laser field pushes the electrons to the right, the total external potential (the sum of the external laser field and the electrostatic potential due to the positron) attracts them to a local minimum to the left. 

For sufficiently weak and sufficiently long (i.e., roughly adiabatically switched-on) laser pulses, the change in the electronic and positronic
position expectation value can be expressed in terms of stationary states $\ket{n}$ and ground-to-excited state transition
energies $\omega_n$ as
\begin{equation}
    \Delta \langle \hat{z} \rangle (t) \approx 2q^{-1} \sum_{n>0} \frac{\omega_n \vert \braket{0 \vert \hat{z} \vert n}\vert^2}{\omega_n^2 - \omega^2}
    \mathcal{E}_0 \sin (\omega t).
\end{equation}
While this linear-response expression suggests that positrons should follow the field and electrons move oppositely in the transparent region $\omega < \omega_1$, it also
indicates that the relative phase angle between the external laser field and the induced dipole moment shifts by $180^\circ$ as the frequency passes through the first transition energy $\omega_1$. This typically happens in ordinary atoms, and we observe the same in the PsA systems. However, both the positron and the electrons go through separate resonant regimes, causing more complex dynamics.

For a closer examination of this phenomenon, we write the induced position expectation value as
\begin{align}
    \Delta \langle \hat{z} \rangle (t) \sim G(t) \sin(\omega t - \theta),
\end{align}
where $G(t)$ varies on a longer time scale than the sine function. This expression gives an approximate definition of the relative phase angle, $\theta$, between the external laser field defined in Eq. \eqref{eq:electric_field} and the induced dipole moment.
In practice, $\langle\hat{z}\rangle$ is obtained from the simulation and we calculate the phase angles approximately following the steps in Appendix \ref{app:phase_angles}.

Figure \ref{fig:phase_angles_PsH} shows the relative phase angle between the maximum laser peak in a weak ($\mathcal{E}_0=0.0001\,\mathrm{a.u.}$) 8-cycle pulse and the corresponding peaks in the dipole moment of the electrons and the positron.
\begin{figure}[h]
\includegraphics[width=8.6cm]{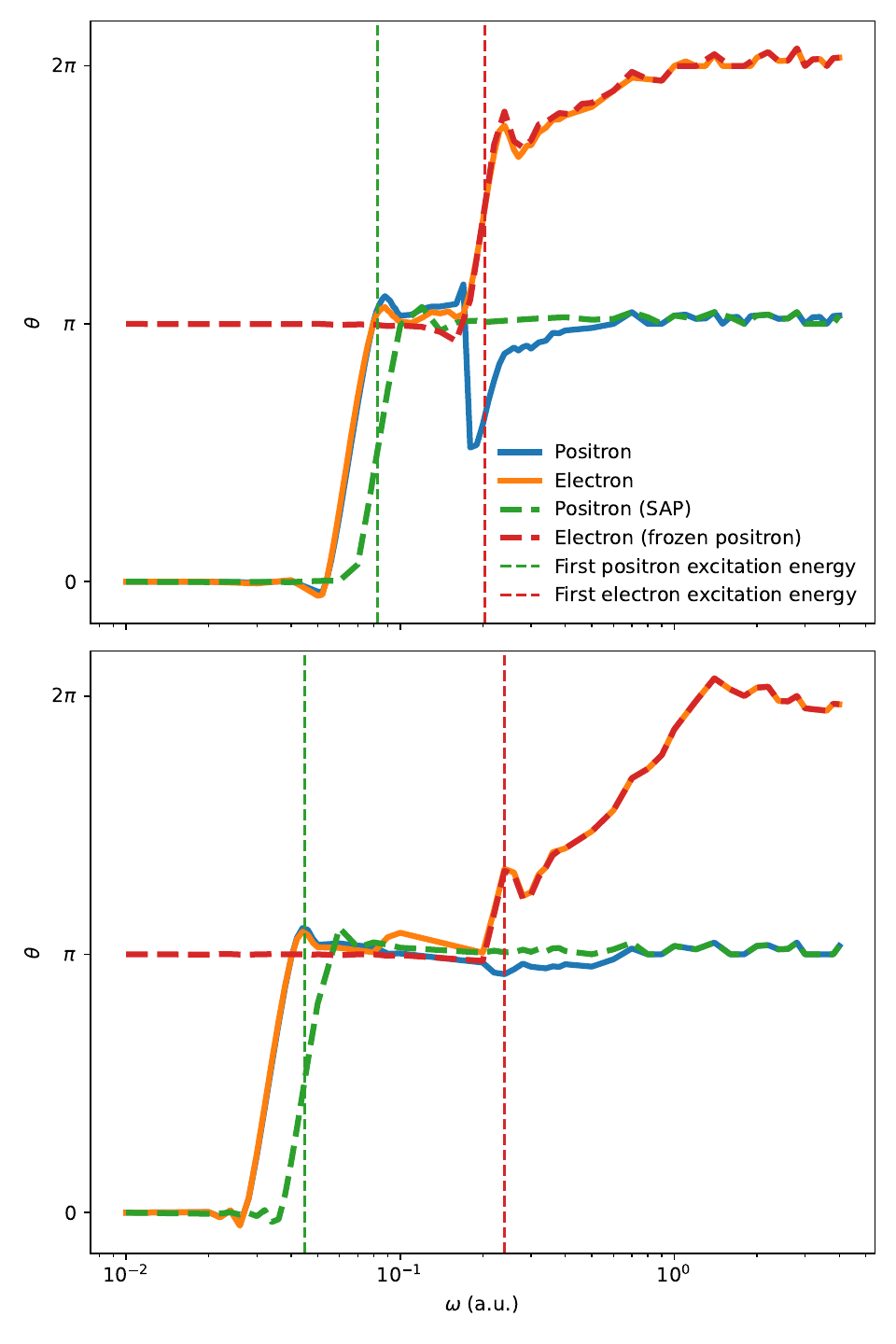}
\caption{\label{fig:phase_angles_PsH} The relative phase angle, $\theta$, between the laser field and $\langle z \rangle$ for the positron and electrons in PsH (top) and PsCl (bottom) as a function of the field frequency. A weak ($\mathcal{E}_0=0.0001\,\mathrm{a.u.}$) 8-cycle laser pulse was applied.}
\end{figure}
This figure also includes the corresponding dipole moments from two reference simulations: the SAP model
and a model in which the positron is kept frozen.

As the frequency approaches the first positronic resonance, the electron and positron oscillation remains in sync.
The dynamics are dominated by the positron, although the electronic dynamics have a small effect on the energy at which the phase shift occurs.
As we reach the first electronic resonance, the roles reverse in the sense that the energy transfer to the electronic system is so large that the electronic dynamics dominate the positron dynamics. This can be seen from the fact that the electronic phase angle is almost identical to the one in the frozen positron simulation, while the positron phase angle deviates substantially from the frozen electron simulation. As seen in the bottom panel of Fig. \ref{fig:dipole_laser_phase}, the magnitude of the dipole moment is substantially greater for the electrons than for the positron in this region.

The discussion so far has been concerned with bound dynamics, but also the ionization dynamics exhibit interesting features.
\begin{figure}[h]
\includegraphics[width=8.6cm]{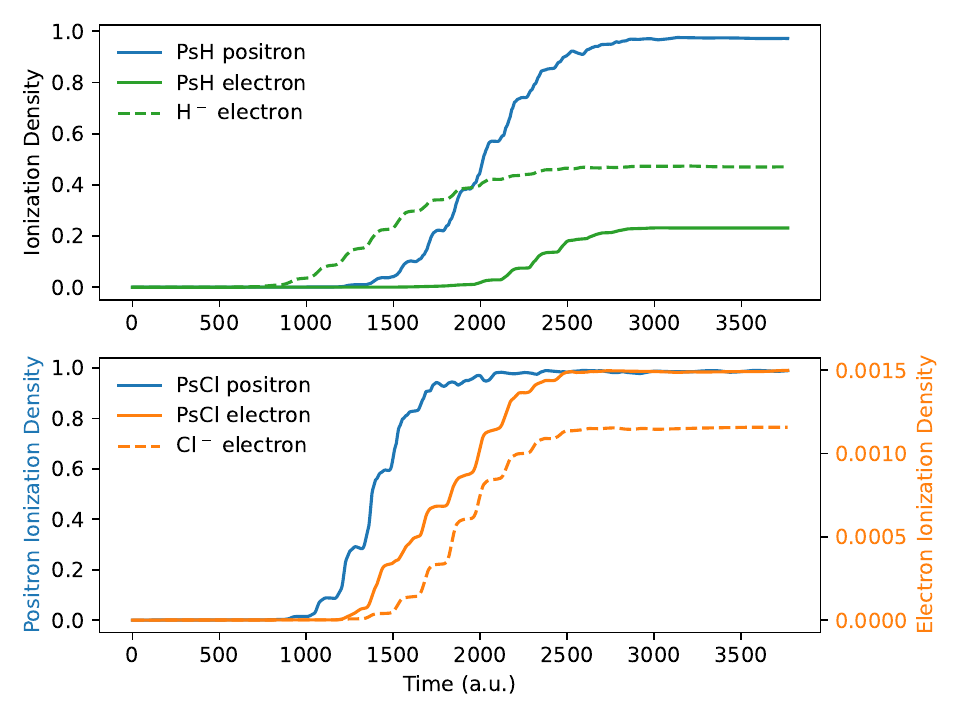}
\caption{\label{fig:ionization_densities} Electron and positron ionization densities for PsH (top, $\mathord{\sim}5.35 \times 10^{12}\ \mathrm{W/cm^2}$) and PsCl (bottom, $\mathord{\sim}4.05 \times 10^{12}\ \mathrm{W/cm^2}$). We used a 12 cycle pulse with frequency 2280 nm. The grid parameters were $r_{\text{max}}=175$ and $l_{\text{max}}=23$, and we used time step $\mathrm{dt}=0.05$ a.u.}
\end{figure}
Figure \ref{fig:ionization_densities} shows the ionization density for the positron and electrons in PsH and PsCl subjected to a laser field with parameters $\omega=0.02\,\mathrm{a.u.}$, and field strengths $\mathcal{E}_0=0.0123443\,\text{a.u.}$ for PsH and H$^-$, and $\mathcal{E}_0=0.01073842\,\text{a.u.}$ for PsCl and Cl$^-$. For PsH, we observe sequential ionization, with the positron ionizing first, followed by the electron. The electron ionization is substantially delayed compared with H$^-$. However, in PsCl the electron ionization is slightly enhanced compared to in Cl$-$ after the initial positron ionization---although, it should be noted that the electron ionization is three orders of magnitude smaller than the positron ionization in this case. It seems as if the positron dynamics induces more electron ionization than the external field itself, consistent with the creation of a small amount of positronium. Correlated wavefunction models would be required to confirm this. The positron-induced electron ionization is further supported by reference simulations holding the positron density fixed in the ground state, which quenches the electron ionization almost completely, to only about $10^{-7}$ at the end of the simulation.

\begin{figure}[h]
\includegraphics[width=8.6cm]{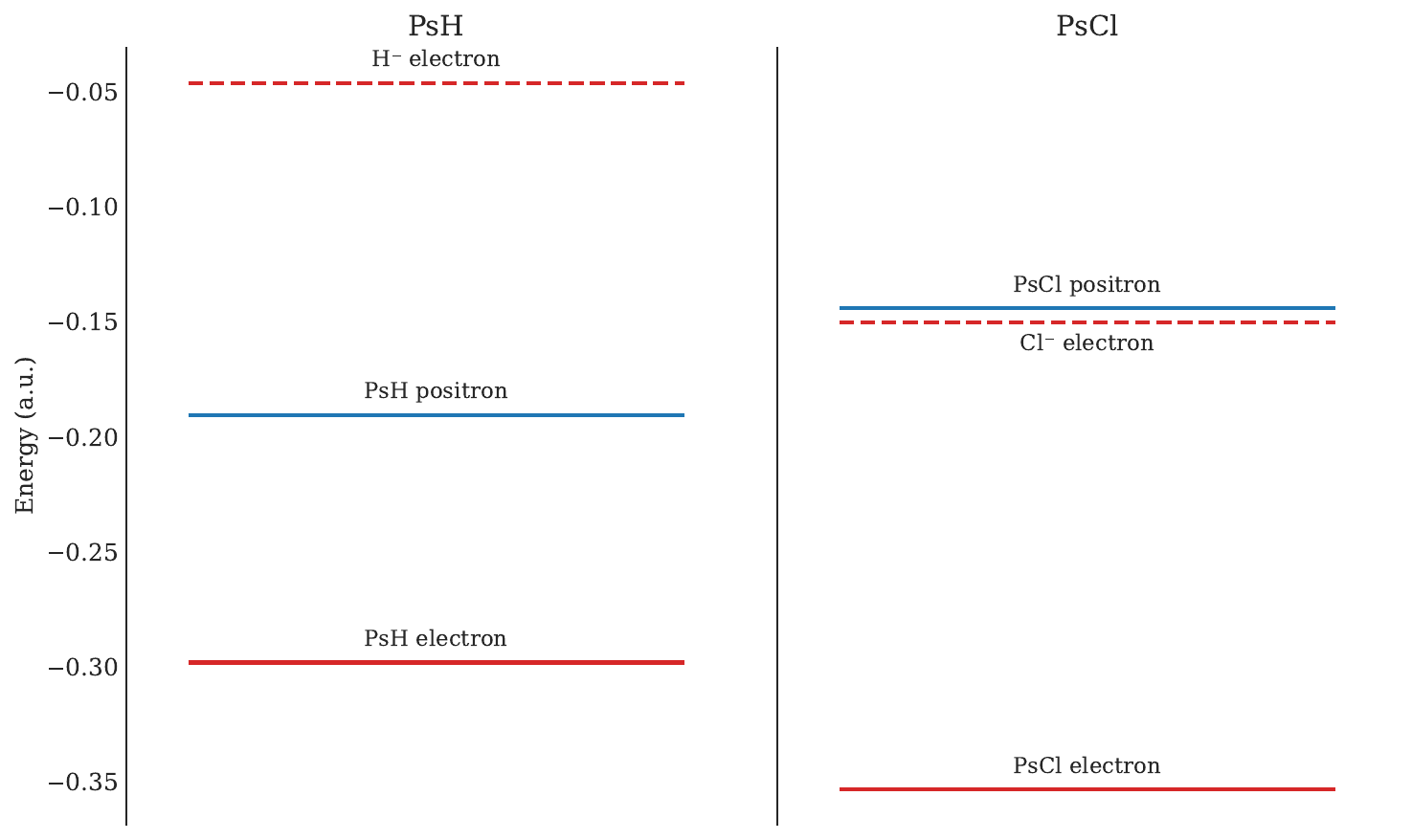}
\caption{\label{fig:orbital_energies} Positron orbital energies and electron valence orbital energies for H$^-$, Cl$^-$, PsH and PsCl.}
\end{figure}

\begin{figure}
\includegraphics[width=8.6cm]{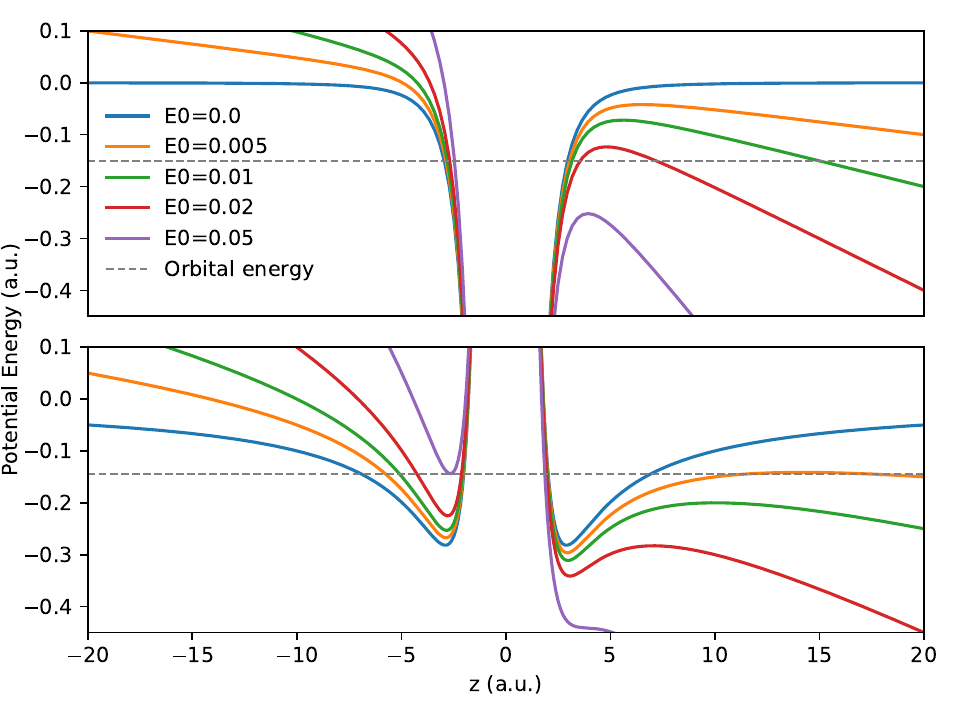}
\caption{\label{fig:1d_potentials_Cl-PsCl} Ground state mean-field potentials along the z-axis seen by the valence electron in Cl$^-$ (top) and the positron in PsCl (bottom) at peak amplitude for various external field strengths. The orbital energy is for the valence electron (top) and positron (bottom).}
\end{figure}

Figure \ref{fig:orbital_energies} shows the difference between the electron and positron energy levels in the two pairs of systems. 
When the positron ionizes in PsH, the ionization potential for the electron is small, while in PsCl, it is about the same as for the positron.
Conspicuously, the ionization of the positron in PsCl is much greater than the electron in Cl$^-$, despite the valence orbital energies being almost the same. This is likely due to the differences in the shape of the effective potential seen by the different particles, as illustrated in Fig. \ref{fig:1d_potentials_Cl-PsCl}.

\subsection{ATI spectra of PsCl and Ps}

\begin{figure}[h]
\includegraphics[width=8.6cm]{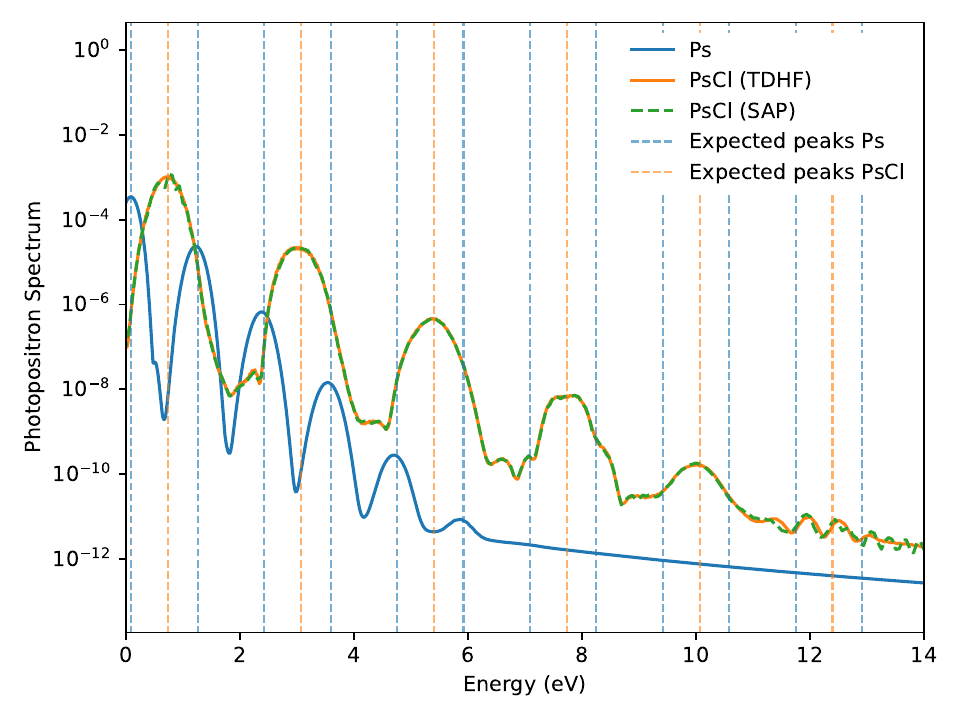}
    \caption{\label{fig:Ps-PsCl_multiphoton_spectrum} Weak-field (10 cycles, $\mathord{\sim}1.66 \times 10^{12}\ \mathrm{W/cm^2}$, $\mathord{\sim}532\ \mathrm{nm}$) photopositron spectrum for Ps and PsCl using the TDHF and SAP models.}
\end{figure}

Figure \ref{fig:Ps-PsCl_multiphoton_spectrum} shows ATI spectra for the positron of Ps and PsCl due to a 10-cycle laser pulse with $\omega=0.085645\,\mathrm{a.u.}$ and $\mathcal{E}_0=0.0068795\,\text{a.u.}$, which corresponds to a laser pulse with wavelength $\lambda \mathord{\sim}532\,\mathrm{nm}$ and intensity $I \mathord{\sim}1.66 \times 10^{12}\,\mathrm{W/cm^2}$. This corresponds to Keldysh parameter values of $\gamma=8.8$ and $\gamma=6.7$ for the Ps and PsCl systems, respectively, placing us firmly within the multiphoton regime for both systems. For the calculation of these spectra, we used $r_{\text{max}}=1000$ and $l_{\text{max}}=L_{\text{max}}=5$, which yields well converged results.
Such a large radial grid is necessary since even the highest-energy (ionized) electrons must not have been absorbed by the mask function at the end of the pulse. We mapped the final state onto a radial grid with $r_{max}=7000$ for the energy analysis, and used a broadening parameter $\Gamma=0.0125 eV$.

Due to the symmetry between the electron and the positron in Ps, the distance between multiphoton ATI peaks is halved relative to an atom. This is because the kinetic energy gained by absorbing a photon is shared equally between the electron and the positron. In a positronium-binding atom, this symmetry is broken. In particular, the electrons are bound tighter around the nucleus in the ground state. The multiphoton ATI patterns can therefore be expected to look more similar to those in an atom. 
This indicates that if PsCl has been successfully produced experimentally, positron ATI peaks should appear at near twice the energy of that in pure Ps.
We see in Fig.~\ref{fig:Ps-PsCl_multiphoton_spectrum} that despite the substantial contributions of the electron potential, the ATI peaks are located as predicted from the positron orbital energy. Moreoever, we note that highest-lying PsCl ATI peaks, above $\mathord{\sim}7\,\mathrm{eV}$, are well separated from the Ps peaks, potentially enabling direct spectroscopic identification of PsCl despite the presence (of relatively large amounts) of Ps. Since the positron binding energy is underestimated at the RHF level, including correlation in the simulation should shift the PsCl peaks somewhat down in energy. It is unlikely, however, that the shift is large enough to place the highest-lying PsCl peaks within the Ps spectrum.

\begin{figure}[h]
\includegraphics[width=8.6cm]{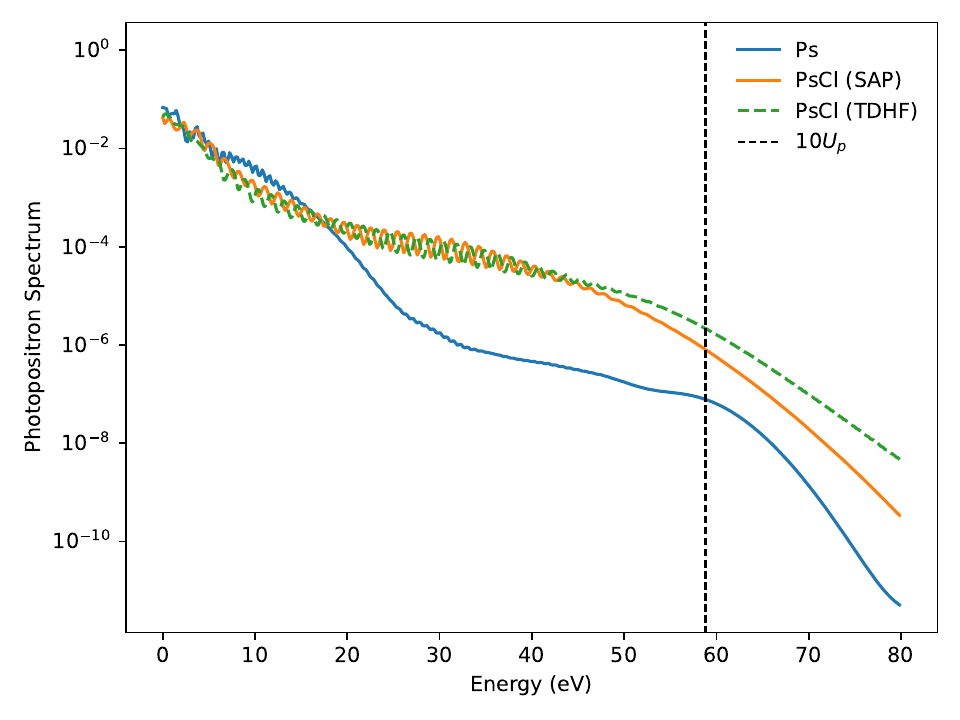}
\caption{\label{fig:Ps-PsCl_rescattering_spectrum} Strong-field (4 cycles, $\mathord{\sim}9.86 \times 10^{13}\,\mathrm{W/cm^2}$, $\mathord{\sim}800\,\mathrm{nm}$) photopositron spectrum for Ps and PsCl using the TDHF and SAP models. Broadening factor was 0.1 eV.}
\end{figure}

The situation becomes somewhat different as we enter the tunneling regime. In this case, a plateau is expected to appear in the ATI spectra due to backscattering.
This process, along with a theoretical cutoff for the maximal energy a particle can acquire, can be largely described classically using the ``simpleman's model''.~\cite{Paulus1994} Energies that ionized electrons can obtain by various processes in different directions can be estimated in terms of the ponderomotive energy, which for an electron or positron is given by
\begin{align}
    U_p = \frac{\mathcal{E}_0^2}{4\omega^2}.
\end{align}
For an atom, the maximal energy the electron can obtain is roughly $10 \,U_p$.
If we replace the electron in the ``simpleman's model'' with an electron-positron pair, the model predicts the maximal energy of each particle in Ps to be $10 \, U_p$, giving the total system $20 \, U_p$ of energy. Thus, the cutoff energy for a positron in Ps is expected to be the same as in PsCl. This is seen in figure \ref{fig:Ps-PsCl_rescattering_spectrum}, which shows ATI spectra for Ps and PsCl after irradiation by a 4-cycle laser pulse with $\omega=0.057\,\mathrm{a.u.}$ ($\mathord{\sim}800\,\mathrm{nm}$) and $\mathcal{E}_0=0.053\,\mathrm{a.u.}$ ($\mathord{\sim}9.86 \times 10^{13}\,\mathrm{W/cm^2}$). These laser parameters correspond to $\gamma=0.76$ and $\gamma=0.57$ for Ps and PsCl, respectively. That is, we are within the tunneling regime.

It is evident from Fig.~\ref{fig:Ps-PsCl_rescattering_spectrum} that the PsCl rescattering should be clearly visible in an experimental spectrum provided that the amount of Ps present is sufficiently low. Although the peaks in the plateau would change, including electron and electron-positron correlations in the simulation is not expected to significantly alter this prediction.

Figures \ref{fig:Ps-PsCl_multiphoton_spectrum} and \ref{fig:Ps-PsCl_rescattering_spectrum} also display the spectra obtained with the SAP model. The SAP model performs very well for PsCl in the multiphoton regime, producing an ATI spectrum virtually identical to that obtained with TDHF theory. Deviations between TDHF and SAP results are only observed at the highest energies from roughly $11\,\mathrm{eV}$ to $14\,\mathrm{eV}$ in Fig.~\ref{fig:Ps-PsCl_multiphoton_spectrum}. In the tunneling regime, however, the SAP model performs significantly worse due to stronger electronic response to the external laser, see Fig.~\ref{fig:Ps-PsCl_rescattering_spectrum}. The cutoff energies observed with the SAP and TDHF models are in reasonable agreement, though.

\section{Concluding remarks}

We have presented a computational study of the coupled electron and positron dynamics driven by laser fields for Ps, PsH, and PsCl at the TDHF level
of theory, using a spherical polar pseudospectral representation to eliminate finite-basis effects and to capture continuum dynamics.
The methodology is described in detail, including an account of the approximate second-order integrator used to solve the TDHF equations of motion for both electronic and positronic degrees of freedom.

Investigating the ``synchronization'' of the electronic and positronic quiver motion along the direction of a weak external electric field observed in a previous work by \citeauthor{Suzuki18},~\cite{Suzuki18} we have argued that it is caused by a shift in the effective potential minimum of the electrons due to the presence of the positron. The effect is frequency-dependent, however, and the relative motion of the electrons and the positron may align or anti-align.

For stronger laser pulses, we find that positron ionization is faster and stronger than electron ionization.
Electron ionization is significantly delayed in PsH relative to the free H$^-$ ion, while a slight enhancement is observed in PsCl relative to Cl$^-$.
These observations may be altered, however, if electron and electron-positron correlations are included in the theoretical model.

We propose that photopositron spectra obtained with short laser pulses at $532\,\mathrm{nm}$ may be used to confirm the
formation of PsCl using the (close to) doubling of the energy range of peaks in the multiphoton ionization regime.
In strong-field cases at $800\,\mathrm{nm}$ (in the tunelling regime), however, the expected plateau in the photopositron spectrum of PsCl will likely only be clearly distinguishable from Ps if its concentration is low enough.
We argue that although the photopositron peaks (position and intensity) will likely change when correlations are included in the theoretical model, these observations will remain valid.

The essential next step to improve and confirm the overall results presented here will be to include electron and electron-positron correlation using the pseudospectral representation.

\appendix

\section{The pseudospectral method} \label{app:pseudospectral_grid}

In this paper, single-particle functions are represented using polar coordinates. A partial wave expansion is employed, leading to a tensor product basis of spherical harmonics and radial basis functions. The radial functions are discretized with a  pseudospectral method we now describe. In the pseudospectral method~\cite{boyd2001chebyshev, cinal2020highly} a function $f(\xi) : [-1,1]\to \mathbb{C}$ is approximated as
\begin{equation}
    f(\xi) \approx f_N(\xi) = \sum_{k=0}^N f(\xi_k) C_k(\xi),
\end{equation}
where $C_k(\xi)$ are the cardinal functions, in our case polynomials of degree $N$ satisfying 
\begin{equation}
    C_k(\xi_j) = \delta_{jk}.
\end{equation}
Thus, we have a global representation in terms of polynomials of degree $N$.  The grid points can be chosen in numerous ways, often as nodes of a Gaussian quadrature rule, leading to highly efficient and accurate evaluation of variational matrix elements and rapid convergence of the polynomial expansion.

In a standard Gauss grid (or ``roots grid''), the grid points (abscissa) are the roots (zeros) of an associated orthogonal polynomial $P_{N}(\xi)$ of order $N$ (Legendre, Chebyshev, Hermite, Laguerre, etc.).

Alternatively, the grid points are taken as the roots ($N-1$) of the derivative of the associated polynomial plus two (specified) endpoints, which is referred to as a Lobatto grid. According to \citeauthor{boyd2001chebyshev}~\cite{boyd2001chebyshev}, the Lobatto grid is favored for solving boundary value problems 
because boundary conditions determine two grid point values of the unknown $f(\xi)$ (or $f_N(\xi)$). Moreover, Lobatto grids can be combined with the finite-element method,~\cite{rescigno2000numerical, bandrauk2011quantum} which can be exploited for increased efficiency and flexibility in the design of the grid.

Integrals over some domain are approximated by an associated Gaussian quadrature rule
\begin{equation}
    \int f(\xi) \mathrm{d}\xi \approx \sum_{k=0}^N f(\xi_k) w_k,
\end{equation}
where $w_k$ are the quadrature weights (see below). 

The derivative of $f(\xi)$ at a grid point $\xi_j$ is approximated by the exact derivative of the polynomial interpolant,
\begin{equation}
    f^\prime(\xi_j) \approx \sum_k f(\xi_k) C_k^\prime(\xi_j) = \sum_k D^{(1)}_{j,k} f(\xi_k), 
\end{equation}
where we have defined the first derivative matrix
\begin{equation}
    D^{(1)} \equiv \left[ D^{(1)}_{j,k} \right] \equiv \left[ C_k^\prime(\xi_j) \right],
\end{equation}
which can be found analytically for certain polynomials such as Legendre and Chebyshev~\cite{boyd2001chebyshev}. 

If the function $f(\xi)$ satisfies homogeneous Dirichlet boundary conditions 
\begin{align}
    f(\xi_0) = 0, \quad f(\xi_N) = 0,
\end{align}
the second derivative matrix for $i,j=1,2,\cdots,N-1$ is given by 
\begin{equation}
    D^{(2)}_{ij} = \sum_{k=1}^{N-1} D^{(1)}_{ik} D^{(1)}_{kj}.
\end{equation}

\subsection*{The Gauss-Legendre-Lobatto grid}

In this work, we have used a Gauss-Legendre-Lobatto (GLL) grid where $P_{N}(\xi), \, \xi \in [-1,1]$ is the Legendre polynomial of order $N$. 
The inner grid points are the roots of $P^\prime_{N}(\xi)$ 
\begin{equation}
    P^\prime(\xi_k) = 0, \qquad k=1,2,\cdots,N,
\end{equation}
and the endpoints are $\xi_0=-1$ and $\xi_{N+1}=1$. The quadrature weights are given by
\begin{equation}
        w_k = \frac{2}{N(N+1)P_N(\xi_k)^2}.
\end{equation}
The grid points and weights can be determined numerically using, for example, as we have done, the \texttt{polynomial}~\cite{numpy_polynomial_legendre} package in \texttt{NumPy}.

Furthermore, the matrix elements of the differentiation operator can be found analytically, see, for example Refs.~\citenum{boyd2001chebyshev, cinal2020highly}~\footnote{the sign for the differentiation matrix at the boundary points $\tilde{C}_0^\prime(\xi_0)$ and $\tilde{C}^\prime_N(\xi_N)$ given in Refs.~\citenum{boyd2001chebyshev, cinal2020highly} is wrong. They should be opposite.}
\begin{align}
    &D^{(1)}_{ij} \equiv \frac{\mathrm{d} C_j}{\mathrm{d}\xi} \Bigr|_{\xi=\xi_i} = C_j^\prime(\xi_i) = \tilde{C}_j^\prime(\xi_i) \frac{P_N(\xi_i)}{P_N(\xi_j)}, \\
    &\tilde{C}^\prime_j(\xi_i) = 
    \begin{cases}
        -\frac{1}{4}N(N+1) & i=j=0, \\
        \phantom{-}\frac{1}{4}N(N+1)  & i=j=N, \\
        \phantom{-}0 & i=j \text{ and } 1 \leq j \leq N-1, \\
        \phantom{-}\frac{1}{\xi_i-\xi_j} & i \neq j.
    \end{cases}
    \label{dg_dx_dg_tilde}
\end{align}

\subsection*{Change of interval}
Since we work in spherical coordinates, the GLL grid points have to be mapped from $[-1,1] \rightarrow [0, r_{\text{max}}]$. We have used the linear mapping
\begin{equation}
    r(\xi) = \frac{r_{\text{max}}}{2} \left( \xi +1\right) \label{eq:gauss_to_rad_linear}.
\end{equation}

Integrals are then given by 
\begin{align}
    \int_0^{r_\text{max}} f(r) \mathrm{d}r &= \int_{-1}^1 f(r(\xi)) \dot{r}(\xi) \mathrm{d}\xi \nonumber \\
    &= \sum_k \dot{r}(\xi_k) f(r(\xi_k)) w_k \label{eq:radial_quad},
\end{align}
where 
\begin{equation}
    \dot{r}(\xi) \equiv \frac{\mathrm{d} r}{\mathrm{d}\xi},
\end{equation}
which for the linear mapping~\eqref{eq:gauss_to_rad_linear} is constant and given by 
\begin{equation}
    \dot{r}(\xi) = \frac{r_{\text{max}}}{2}.
\end{equation}

We stress that, as discussed in section~\ref{sec:model_and_computational_method}, the radial components of the direct and exchange potentials defined by Eqs.~\eqref{eq:radial_dir_exc_pot_electrons} and \eqref{eq:radial_dir_pot_positron} are \textit{not} computed by the quadrature formula~\eqref{eq:radial_quad}, but are instead found by solving an associated Poisson equation.

To express the time-dependent radial TDHF equations of motion~\eqref{eq:radial_eom_electron_orbs} and~\eqref{eq:radial_eom_positron_orbs} in the pseudospectral basis one needs to relate the first and second derivative of a function $f(r)$ with respect to $r$ to the variable $\xi$. By the chain rule we obtain
\begin{align}
    \frac{\mathrm{d} f(r)}{\mathrm{d} r} &= \frac{1}{\dot{r}(\xi)} \frac{\mathrm{d} f}{\mathrm{d}\xi}, \\
    \frac{\mathrm{d}^2 f(r)}{\mathrm{d} r^2} &= \frac{1}{\dot{r}(\xi)^2} \frac{\mathrm{d}^2 f}{\mathrm{d}\xi^2} - \frac{\ddot{r}(\xi)}{\dot{r}(\xi)^3} \frac{\mathrm{d} f}{\mathrm{d}\xi},
\end{align}
where the second term in the double derivative vanishes for the linear mapping since $\ddot{r}(\xi) = 0$.

\section{Time integration}
\label{app:propagator}
\subsection{The TDHF equations on matrix form}

We now indicate how the TDHF equations are written as a \textit{non-linear} matrix equation. Here, we consider the electronic (restricted) TDHF equation for the occupied orbitals ($i=1,\cdots,N_e/2$) given by
\begin{equation}
    \mathrm{i} \ket{\dot{\phi}_i(t)} = \left( \hat{h}(t) + \hat{v}^{\text{RHF}} \right) \ket{\phi(t)} \label{eq:tdhf_ket_form},
\end{equation}
where 
\begin{equation}
    \hat{v}^{\text{RHF}} \ket{\phi_i} = 2 \sum_{j=1}^{N_e/2} \braket{\cdot \phi_j|\phi_i \phi_j} - \sum_{j=1}^{N_e/2} \braket{\cdot \phi_j|\phi_j \phi_i}. 
\end{equation}
Here, we have defined
\begin{equation}
    \braket{\cdot \phi_j|\phi_k \phi_l} \equiv \left(\int \frac{\phi^*_j(\mathbf{r}_2) \phi_l(\mathbf{r}_2)}{|\mathbf{r}_1-\mathbf{r}_2|} \mathrm{d}\mathbf{r}_2\right) \ket{\phi_k}, 
\end{equation}
such that 
\begin{align}
    &\bra{\chi}\braket{\cdot \phi_j|\phi_k \phi_l} \equiv \braket{\chi \phi_j|\phi_k \phi_l} \\ 
    &= \iint \frac{\chi^*(\mathbf{r}_1)\phi^*_j(\mathbf{r}_2) \phi_k(\mathbf{r}_1)\phi_l(\mathbf{r}_2)}{|\mathbf{r}_1-\mathbf{r}_2|} \mathrm{d}\mathbf{r}_1 \mathrm{d}\mathbf{r}_2.
\end{align}

The formulation of the full set of equations for electrons and a positron is analogous.

Now, assume that the time-dependent orbitals are expanded in a finite orthonormal basis
\begin{equation}
    \ket{\phi_i(t)} = \sum_{\beta=1}^{N_b} C_{\beta i}(t) \ket{\chi_\beta}, \qquad \braket{\chi_\alpha|\chi_\beta} = \delta_{\alpha \beta},
\end{equation}
where $N_b$ denotes the number of basis functions. In the present case, $\chi_\alpha(\mathbf{r})$ is a tensor product of a spherical harmonic and a cardinal function.
Projecting Eq.~\eqref{eq:tdhf_ket_form} with $\bra{\chi_\alpha}$ we find that 
\begin{align}
    \mathrm{i} \dot{C}_{\alpha,i}(t) &= \sum_\gamma h_{\alpha \gamma}(t) C_{\gamma,i}(t) + 2 \sum_{j=1}^{N_e/2} \braket{\chi_\alpha \phi_j|\phi_i \phi_j}  \nonumber \\
    &- \sum_{j=1}^{N_e/2} \braket{\chi_\alpha \phi_j|\phi_j \phi_i} \\
    &= \sum_\gamma h_{\alpha \gamma}(t) C_{\gamma,i}(t) + 2 \sum_{\gamma} V^{\text{dir}}_{\alpha \gamma}(C(t)) C_{\gamma,i}(t) \nonumber \\
    &-\sum_{\gamma} V^{\text{exc}}_{\alpha \gamma}(C(t)) C_{\gamma,i}(t) \\ &= \sum_\gamma F_{\alpha \gamma}(t, C(t)) C_{\gamma, i}(t) \label{eq:tdhf_basis_component form},
\end{align}
where we have defined the one-body matrix elements,
\begin{align}
    h_{\alpha \gamma}(t) &\equiv \braket{\chi_\alpha|\hat{h}(t)|\chi_\gamma},
\end{align}
the direct and exchange matrices,
\begin{align}
    V^{\text{dir}}_{\alpha \gamma}(C(t)) &\equiv \sum_{\beta \delta} \rho_{\beta \delta}(C(t)) u^{\alpha \beta}_{\gamma \delta} \label{eq:dir_matrix}\\
    V^{\text{exc}}_{\alpha \gamma}(C(t)) &\equiv \sum_{\beta \delta} \rho_{\beta, \delta}(C(t)) u^{\alpha \beta}_{\delta \gamma } \label{eq:exc_matrix} \\
    u^{\alpha \beta}_{\gamma \delta} &\equiv \braket{\chi_\alpha \chi_\beta|\chi_\gamma \chi_\delta},
\end{align}
the time-dependent Hartree-Fock density matrix,
\begin{align}
    \rho_{\beta \delta}(C(t)) &\equiv \sum_{j=1}^{n/2} C^*_{\beta j}(t)C_{\delta j}(t) \label{eq:hf_density_def},
\end{align}
and the time-dependent Fock matrix,
\begin{equation}
    F_{\alpha \gamma}(t,C(t)) \equiv h_{\alpha \gamma}(t)+ 2 V^{\text{dir}}_{\alpha \gamma}(C(t)) - V^{\text{exc}}_{\alpha \gamma}(C(t)).
\end{equation}

Eq.~\eqref{eq:tdhf_basis_component form} can be written on matrix form as 
\begin{equation}
    \mathrm{i} \dot{C}(t) = F(t, C(t)) C(t) \label{eq:tdhf_roothan_hall},
\end{equation}
where 
\begin{align}
    C(t) &\equiv [C_{\gamma i}(t)] \in \mathbb{C}^{N_b \times n/2}, \\
    F(t,C(t)) &\equiv [F_{\alpha \gamma}(t,C(t))] \in \mathbb{C}^{N_b \times N_b}.
\end{align}
We emphasize that Eq.~\eqref{eq:tdhf_roothan_hall} is a non-linear ordinary differential equation in the sense that the matrix $F(t,C(t))$ is dependent on the unknowns $C(t)$.

\subsection{Implicit midpoint/Crank-Nicolson based constant density matrix schemes}

In order to integrate Eq.~\eqref{eq:tdhf_roothan_hall} wtih a given initial condition, we first observe that the presence of $\nabla^2$ in the Fock operator leads to $F(t,C(t))$ having eigenvalues that grow quadratically with the number of radial grid points. Explicit Runge--Kutta methods would lead to excessively stringent conditions on the time step. A straightforward implicit method that is symplectic and in particular norm-preserving  \cite{hairerGeometricNumericalIntegration2006} is the implicit midpoint method, yielding
\begin{align}
    C^{k+1} = C^k - \frac{\mathrm{i} \Delta t}{2} F(t_{k+\frac{1}{2}}, \bar{C}) \left( C^k + C^{k+1} \right) \label{eq:tdhf_implicit_midpoint},
\end{align}
where for a chosen time step $\Delta t$,
\begin{align}
    C^k &\equiv C(t_k), \quad t_k = k\Delta t, \quad k=0,1,\cdots, k_{\text{steps}}, \\
    t_{k+\frac{1}{2}} &\equiv t_k + \frac{\Delta t}{2}, \\
    \bar{C} &\equiv \frac{1}{2} \left( C^k + C^{k+1} \right).
\end{align}
Re-arranging terms, Eq.~\eqref{eq:tdhf_implicit_midpoint} can be written as
\begin{equation}
    \left(I + \frac{\mathrm{i} \Delta t}{2} F(t_{k+\frac{1}{2}}, \bar{C}) \right) C^{k+1} = \left(I - \frac{\mathrm{i} \Delta t}{2} F(t_{k+\frac{1}{2}}, \bar{C}) \right) C^k \label{eq:tdhf_crank_nicolson_form},
\end{equation}
which is a system of non-linear equations because $F(\bar{t}, \bar{C})$ is dependent on the unknown $C^{k+1}$ through $\bar{C}$. However, for a fixed $\bar{C}$, it is a system of \emph{linear} equations.

To solve Eq.~\eqref{eq:tdhf_crank_nicolson_form}, we approximate $\bar{C}$. 
As a first approximation, we take 
\begin{equation}
     \bar{C} = \frac{1}{2} \left( C^k + C^{k+1} \right) \approx C^k \label{eq:barC_approx},
\end{equation}
which can be interpreted as keeping the density matrix~\eqref{eq:hf_density_def} (and consequently the direct~\eqref{eq:dir_matrix} and exchange~\eqref{eq:exc_matrix} matrices) constant over the interval $[t, t+\Delta t]$, which we refer to as constant density-matrix (CDM) integration. Solving Eq.~\eqref{eq:tdhf_implicit_midpoint} with $\bar{C}$ given by Eq.~\eqref{eq:barC_approx} yields an integration scheme that is accurate to first order in $\Delta t$, which we refer to as the CDM1 integrator.

An approximate second-order scheme, the CDM2 integrator, can be formulated using an idea of \citeauthor{beck1997efficient}~\cite{beck1997efficient} The CDM2 integrator can be described as follows:
\begin{itemize}
    \item [(i)] Solve Eq.~\eqref{eq:tdhf_implicit_midpoint} a \textit{half step} forwards in time (from $[t, t+\frac{\Delta t}{2}]$) using Eq.~\eqref{eq:barC_approx} for $\bar{C}$, i.e., solve 
    \begin{align}
        &\left(I + \frac{\mathrm{i} \Delta t}{4} F(t_{k+\frac{1}{4}}, \bar{C}) \right) \tilde{C} = \nonumber \\
        &\left(I - \frac{\mathrm{i} \Delta t}{4} F(t_{k+\frac{1}{4}}, \bar{C}) \right) C^k, \label{eq:step1_CDM2}
    \end{align}
    for $\tilde{C}$.
    \item [(ii)] Use the solution of Eq.~\eqref{eq:step1_CDM2} as an improved estimate for $\Bar{C} \leftarrow \tilde{C}$.
    \item [(iii)] Solve for $C^{k+1}$ in two \textit{half} steps:
    \begin{itemize}
        \item [(a)] First, solve for the midpoint $C^{k+\frac{1}{2}}$
            \begin{align}
                &\left(I + \frac{\mathrm{i} \Delta t}{4} F(t_{k+\frac{1}{4}}, \tilde{C}) \right) C^{k+\frac{1}{2}} = \nonumber \\
                &\left(I - \frac{\mathrm{i} \Delta t}{4} F(t_{k+\frac{1}{4}}, \tilde{C}) \right) C^k
            \end{align}
        \item [(b)] Second, solve for $C^{k+1}$
            \begin{align}
                &\left(I + \frac{\mathrm{i} \Delta t}{4} F(t_{k+\frac{3}{4}}, \tilde{C}) \right) C^{k+1} = \nonumber \\
                &\left(I - \frac{\mathrm{i} \Delta t}{4} F(t_{k+\frac{3}{4}}, \tilde{C}) \right) C^{k+\frac{1}{2}}.
            \end{align}
    \end{itemize}
\end{itemize}
In practice, we have experienced that the CDM2 integrator provides significantly improved accuracy over CDM1 in the sense
that results converge faster with respect to $\Delta t$.

The CDM1 and CDM2 integrators are applicable to the solution of the radial equations of motion for for electrons and a (single) positron given by Eqs.~\eqref{eq:radial_eom_electron_orbs} and \eqref{eq:radial_eom_positron_orbs}.
Applying the implicit midpoint rule, the equations of motion for the electrons and the positron can be written as
\begin{align}
    \left(I + \frac{\mathrm{i} \Delta t}{2} \mathcal{F}_e(t_{k+\frac{1}{2}}; \bar{u}, \bar{v}) \right) u^{k+1} &= \left(I - \frac{\mathrm{i} \Delta t}{2} \mathcal{F}_e(t_{k+\frac{1}{2}}; \bar{u}, \bar{v}) \right) u^{k}, \label{eq:cn_electrons_spherical} \\
    \left(I + \frac{\mathrm{i} \Delta t}{2} \mathcal{F}_p(t_{k+\frac{1}{2}}; \bar{u}) \right) v^{k+1} &= \left(I - \frac{\mathrm{i} \Delta t}{2} \mathcal{F}_p(t_{k+\frac{1}{2}}; \bar{u}) \right) v^{k} \label{eq:cn_positron_spherical}
\end{align}
where $u^n$ and $v^n$ are the vectorization of the discretized electron and positron orbitals
\begin{align}
    u^k &\equiv [u_0^{1,k},\cdots,u_l^{1,k},\cdots, u_l^{N_e/2,k}]^T, \\
    v^k &\equiv [v_0^{1,k},\cdots,v_l^{1,k}]^T
\end{align}
and where $\mathcal{F}_e \in \mathbb{C}^{\left( (N_e/2) N_r n_{lm} \right) \times \left( (N_e/2) N_r n_{lm} \right)}$ and $\mathcal{F}_p \in \mathbb{C}^{( N_r n_{lm}) \times (N_r n_{lm})}$ are matrices that act on the vectorization of the orbitals, $N_r$ denotes the number of radial grid points and $n_{lm}$ the number of angular momenta used in the expansion of the orbital. 

The dimensionality of $\mathcal{F}_e$ and $\mathcal{F}_p$ quickly prohibits the direct solution Eqs.~\eqref{eq:cn_electrons_spherical} and~\eqref{eq:cn_positron_spherical}. 
Instead, we solve the linear equations iteratively with the BiConjugate Gradient Stabilized method (BiCGSTAB)~\cite{barrett1994templates} (implemented in \texttt{SciPy}), which requires the action of $\mathcal{F}_e$ and $\mathcal{F}_p$ on an orbital vector, which for the electrons is given by
\begin{align}
    &(\mathcal{F}_e(t_{k+\frac{1}{2}}; \bar{u}, \bar{v})u^k)^i = \hat{h}^e_l u_{l}^{i,k} \nonumber \\
    &-\mathrm{i}A(t_{k+\frac{1}{2}}) [ a_{l-1, m_i}D_{-l}u_{l-1}^{i,k} + a_{l, m_i}D_{l+1} u_{l+1}^{i,k} ]\nonumber \\
    &+2\sum_{l'} \mathfrak{D}_{l,l'}^{m_i}(r; \bar{u}) u_{l'}^{i,k} -\sum_{l'} \mathfrak{P}_{l,l'}^{m_i}(r; \bar{v}) u_{l'}^{i,k} \nonumber \\
    &- \sum_{j} \sum_{l'} \mathfrak{X}_{ij, ll'}^{m_im_j}(r; \bar{u}_j, u_i^k) \bar{u}_{l'}^{j,k}
\end{align}
where the dependence of the exchange matrix $\mathfrak{X}$ on both $\bar{u}$ and $u^k$ should be noted. For the positron
\begin{align}
    &(\mathcal{F}_p(t_{k+\frac{1}{2}}; \bar{u})v^k) =\hat{h}^p_l v^k_{l} \nonumber \\
    &+\mathrm{i}A(t_{k+\frac{1}{2}}) [ a_{l-1, 0}D_{-l}v^k_{l-1} + a_{l, 0}D_{l+1} v^k_{l+1} ]\nonumber \\
    & -2\sum_{l'} \mathfrak{D}_{l,l'}^{0}(r, \bar{u}) v^k_{l'}.
\end{align}
Additionally, we have found that using a preconditioner is crucial when employing the BiCGSTAB routine. The preconditioners are chosen to approximate $\left(I + \frac{\mathrm{i} \Delta t}{2} \mathcal{F}_e(t_{k+\frac{1}{2}}; \bar{u}, \bar{v}) \right)^{-1}$ and $\left(I + \frac{\mathrm{i} \Delta t}{2} \mathcal{F}_p(t_{k+\frac{1}{2}}; \bar{u}) \right)^{-1}$. 
Exploiting that the Laplacian (and thus the kinetic energy matrix) is block-diagonal in the angular-momentum quantum number, and that the Coulomb interaction of the electrons and the positron with the nucleus is diagonal and time-independent, one can precompute the preconditioner matrices $\mathcal{M}_e$ and $\mathcal{M}_p$ prior to propagation. The preconditioner matrices are thus block-diagonal in $l$, with each $l$-block given by
\begin{align}
    \mathcal{M}^e_l &= \left(I + \frac{i \Delta t}{2} H^e_l  \right)^{-1}, \\
    \mathcal{M}^p_l &= \left(I + \frac{i \Delta t}{2} H^p_l \right)^{-1},
\end{align}
where 
\begin{align}
    H^e_l &= T_l - V, \\
    H^p_l &= T_l + V.
\end{align}
Here, $T_l$ is the matrix representation of $-\frac{1}{2}\Delta_l$ and $V$ is the diagonal matrix representation of the Coulomb interaction $V = [\frac{1}{r_\alpha} \delta_{\alpha \beta}]$ in the GLL basis.

\section{Phase angles}\label{app:phase_angles}

The relative phase angles between the external electric field and the induced position expectation values of the electrons and positrons are calculated by the following steps:
\begin{enumerate}
    \item At $\omega=0.01$, we use $t_\mathrm{peak}$ to indicate the time at which the displacements of the positron and electron peaks. 
        Due to the linear and adiabatic response at this frequency, $t_\mathrm{peak}$ also indicates the time of peak electric-field strength.
    \item By making the $\omega$-steps sufficiently small, we can follow how $t_{\text{peak}}$ evolves as a function of $\omega$.
    \item We then define the phase angle as
    \begin{align}
        \theta = \pi \frac{ t_\mathrm{peak} - t_\mathrm{max} }{|t_\mathrm{min} - t_\mathrm{max}|}, 
    \end{align}
    where $t_{\mathrm{max}} = \underset{t}{\operatorname{argmax}}\, \mathcal{E}(t)$ and $t_{\mathrm{min}} = \underset{t}{\operatorname{argmin}}\, \mathcal{E}(t)$.
\end{enumerate}

\bibliography{refs}

@misc{numpy_polynomial_legendre,
  title = {Numpy polynomial Legendre},
  url = {https://numpy.org/doc/stable/reference/routines.polynomials.legendre.html},
  note = {Accessed on December 19, 2025},
  year = {2025}
}

@book{barrett1994templates,
  title={{Templates for the Solution of Linear Systems: Building Blocks for Iterative Methods}},
  author={Barrett, Richard and Berry, Michael and Chan, Tony F and Demmel, James and Donato, June and Dongarra, Jack and Eijkhout, Victor and Pozo, Roldan and Romine, Charles and Van der Vorst, Henk},
  year={1994},
  publisher={SIAM},
  address = {Philadelphia},
  doi = {10.1137/1.9781611971538}
}

@article{sato2016time,
  title={Time-dependent complete-active-space self-consistent-field method for atoms: Application to high-order harmonic generation},
  author={Sato, Takeshi and Ishikawa, Kenichi L and B{\v{r}}ezinov{\'a}, Iva and Lackner, Fabian and Nagele, Stefan and Burgd{\"o}rfer, Joachim},
  journal={Physical Review A},
  volume={94},
  number={2},
  pages={023405},
  year={2016},
  publisher={APS}
}

@book{hairerGeometricNumericalIntegration2006,
  title = {{Geometric Numerical Integration}},
  author = {Hairer, E. and Lubich, C. and Wanner, G.},
  year = {2006},
  publisher = {Springer},
  address = {Berlin},
  doi = {10.1007/3-540-30666-8}
}

@article{rescigno2000numerical,
  title={Numerical grid methods for quantum-mechanical scattering problems},
  author={Rescigno, Thomas N and McCurdy, C William},
  journal={Phys. Rev. A},
  volume={62},
  number={3},
  pages={032706},
  year={2000},
  publisher={APS}
}

@book{bandrauk2011quantum,
  title={{Quantum Dynamic Imaging: Theoretical and Numerical Methods}},
  author={Bandrauk, Andr{\'e} D and Ivanov, Misha},
  year={2011},
  publisher={Springer},
  address = {New York},
  doi = {10.1007/978-1-4419-9491-2}
}

@article{cade1977electronic,
  title={{The electronic structure and positron annihilation characteristics of positronium halides, [X$^-$; e$^+$]. I. Hartree--Fock calculations and stability}},
  author={Cade, Paul E and Farazdel, Abbas},
  journal={J. Chem. Phys.},
  volume={66},
  pages={2598--2611},
  year={1977}
}

@article{beck1997efficient,
  title={{An efficient and robust integration scheme for the equations of motion of the multiconfiguration time-dependent Hartree (MCTDH) method}},
  author={Beck, Michael H and Meyer, H-D},
  journal={Z. Phys. D},
  volume={42},
  pages={113--129},
  year={1997}
}

@article{kurtz1980theoretical,
  title={Theoretical studies of positron complexes with atomic anions},
  author={Kurtz, Henry A and Jordan, Kenneth D},
  journal={J. Chem. Phys.},
  volume={72},
  pages={493--503},
  year={1980}
}

@article{Leveque24,
    author = {Lévêque-Simon, K. and Camper, A. and Taïeb, R. and Caillat, J. and Lévêque, C. and Giner, E.},
    title = {{Production of positronium chloride: A study of the charge exchange reaction between Ps and Cl$^-$}},
    journal = {J. Chem. Phys.},
    volume = {160},
    pages = {104301},
    year = {2024},
    doi = {10.1063/5.0182498}
}

@article{Fittinghoff92,
  title = {Observation of nonsequential double ionization of helium with optical tunneling},
  author = {Fittinghoff, D. N. and Bolton, P. R. and Chang, B. and Kulander, K. C.},
  journal = {Phys. Rev. Lett.},
  volume = {69},
  pages = {2642--2645},
  year = {1992},
  doi = {10.1103/PhysRevLett.69.2642}
}

@article{Blaga12,
    author = {Blaga, C. I. and Xu, J. and DiChiara, A. D. and Sistrunk, E. and Zhang, K. and Agostini, P. and Miller, T. A. and DiMauro, L. F. and Lin, C. D.},
    year = {2012},
    title ={Imaging ultrafast molecular dynamics with laser-induced electron diffraction},
    journal = {Nature},
    volume = {483},
    pages = {194--197},
    doi = {10.1038/nature10820},
}

@Article{Charry22,
author ="Charry, Jorge and Moncada, Félix and Barborini, Matteo and Pedraza-González, Laura and Varella, Márcio T. do N. and Tkatchenko, Alexandre and Reyes, Andrés",
title  ="The three-center two-positron bond",
journal  ="Chem. Sci.",
year  ="2022",
volume  ="13",
pages  ="13795--13802",
doi  ="10.1039/D2SC04630J"
}

@article{Charry18,
author = {Charry, Jorge and Varella, Márcio T. do N. and Reyes, Andrés},
title = {{Binding Matter with Antimatter: The Covalent Positron Bond}},
journal = {Angew. Chem. Int. Ed.},
volume = {57},
pages = {8859--8864},
doi = {10.1002/anie.201800914},
year = {2018}
}

@article{Gribakin02,
title = {{Enhancement of positron annihilation on molecules due to vibrational Feshbach resonances}},
journal = {Nuc. Inst. Meth. Phys. Res. Sec. B},
volume = {192},
number = {1},
pages = {26--39},
year = {2002},
doi = {10.1016/S0168-583X(02)00702-4},
author = {G.F. Gribakin},
}

@book{boyd2001chebyshev,
  title={{Chebyshev and Fourier spectral methods}},
  author={Boyd, John P},
  year={2001},
  publisher={Dover},
  Address={New York}
}

@book{bransden2003physics,
  title={Physics of atoms and molecules},
  author={Bransden, Brian Harold and Joachain, Charles Jean},
  year={2003},
  edition={2nd},
  publisher={Prentice Hall},
  address={Harlow, England}
}

@article{hochstuhl2014time,
  title={Time-dependent multiconfiguration methods for the numerical simulation of photoionization processes of many-electron atoms},
  author={Hochstuhl, David and Hinz, Christopher M and Bonitz, Michael},
  journal={Eur. Phys. J. Spec. Top.},
  volume={223},
  number={2},
  pages={177--336},
  year={2014},
  publisher={Springer}
}

@article{cinal2020highly,
  title={{Highly accurate numerical solution of Hartree--Fock equation with pseudospectral method for closed-shell atoms}},
  author={Cinal, M},
  journal={J. Math. Chem.},
  volume={58},
  number={8},
  pages={1571--1600},
  year={2020},
  publisher={Springer}
}

@article{Archila24,
author = {Archila, David and Moncada, Felix and Charry, Jorge and Varella, Marcio and Flores-Moreno, Roberto and Torres, Fernando Javier and Reyes, Andrés},
title = {{Two-positron-bonded dihalides: Ps2XY (X,Y=F,Cl,Br)}},
journal = {Chem. Eur. J.},
volume = {30},
pages = {e202402618},
year = {2024},
doi = {10.1002/chem.202402618},
}

@article{Merritt21,
author = {Merritt, Isabella C. D. and Jacquemin, Denis and Vacher, Morgane},
title = {{Attochemistry: Is Controlling Electrons the Future of Photochemistry?}},
journal = {J. Phys. Chem. Lett.},
volume = {12},
pages = {8404--8415},
year = {2021},
doi = {10.1021/acs.jpclett.1c02016}
}

@article{Hofierka22,
    author = {Hofierka, Jaroslav and  Cunningham, Brian  Rawlins, Charlie M. and  Patterson, Charles H. and Green, Dermot G.},
    year = {2022},
    title = {Many-body theory of positron binding to polyatomic molecules},
    journal = {Nature},
    pages = {688--693},
    volume = {606},
    issue = {7915},
    doi =  {10.1038/s41586-022-04703-3}
}

@article{charry_martinez_correlated_2022,
    title = {{Correlated Wave Functions for Electron–Positron Interactions in Atoms and Molecules}},
    volume = {18},
    doi = {10.1021/acs.jctc.1c01193},
    journal = {J. Chem. Theory Comput.},
    author = {Charry Martinez, Jorge Alfonso and Barborini, Matteo and Tkatchenko, Alexandre},
    year = {2022},
    pages = {2267--2280}
}

@article{upadhyay_capturing_2024,
    title = {{Capturing Correlation Effects in Positron Binding to Atoms and Molecules}},
    volume = {20},
    doi = {10.1021/acs.jctc.4c00727},
    journal = {J. Chem. Theory Comput.},
    author = {Upadhyay, Shiv and Benali, Anouar and Jordan, Kenneth D.},
    year = {2024},
    pages = {9879--9893}
}

@article{tumakov_relativistic_2024,
    title = {Relativistic corrections in the ground and excited states of positronic beryllium},
    volume = {109},
    doi = {10.1103/PhysRevA.109.042826},
    journal = {Phys. Rev. A},
    author = {Tumakov, Dmitry and Rzhevskii, Pavel and Shomenov, Toreniyaz and Bubin, Sergiy},
    year = {2024},
    pages = {042826}
}

@article{moncada_covalent_2019,
    title = {Covalent bonds in positron dihalides},
    volume = {11},
    doi = {10.1039/C9SC04433G},
    journal = {Chem. Sci.},
    author = {Moncada, Félix and Pedraza-González, Laura and Charry, Jorge and Varella, Márcio T. do N. and Reyes, Andrés},
    year = {2019},
    pages = {44--52}
}

@article{Ferray88,
doi = {10.1088/0953-4075/21/3/001},
year = {1988},
volume = {21},
pages = {L31},
author = {M Ferray and A L'Huillier and X F Li and L A Lompre and G Mainfray and C Manus},
title = {Multiple-harmonic conversion of 1064 nm radiation in rare gases},
journal = {J. Phys. B}
}

@article{McPherson87,
author = {A. McPherson and G. Gibson and H. Jara and U. Johann and T. S. Luk and I. A. McIntyre and K. Boyer and C. K. Rhodes},
journal = {J. Opt. Soc. Am. B},
pages = {595--601},
title = {Studies of multiphoton production of vacuum-ultraviolet radiation in the rare gases},
volume = {4},
year = {1987},
doi = {10.1364/JOSAB.4.000595}
}

@article{Deng16,
  title = {Ultrafast Excitation of an Inner-Shell Electron by Laser-Induced Electron Recollision},
  author = {Deng, Yunpei and Zeng, Zhinan and Jia, Zhengmao and Komm, Pavel and Zheng, Yinhui and Ge, Xiaochun and Li, Ruxin and Marcus, Gilad},
  journal = {Phys. Rev. Lett.},
  volume = {116},
  pages = {073901},
  year = {2016},
  doi = {10.1103/PhysRevLett.116.073901}
}

@article{Lewenstein94,
  title = {Theory of high-harmonic generation by low-frequency laser fields},
  author = {Lewenstein, M. and Balcou, Ph. and Ivanov, M. Yu. and L'Huillier, Anne and Corkum, P. B.},
  journal = {Phys. Rev. A},
  volume = {49},
  issue = {3},
  pages = {2117--2132},
  year = {1994},
  doi = {10.1103/PhysRevA.49.2117}
}

@article{Muller08,
  title = {Muon pair creation from positronium in a linearly polarized laser field},
  author = {M\"uller, Carsten and Hatsagortsyan, Karen Z. and Keitel, Christoph H.},
  journal = {Phys. Rev. A},
  volume = {78},
  pages = {033408},
  year = {2008},
  doi = {10.1103/PhysRevA.78.033408}
}

@article{Ditmire99,
    author = {Ditmire, T. and Zweiback, J. and Yanovsky, V. P. and Cowan, T. E. and Hays, G. and Wharton, K. B.},
    title = {Nuclear fusion from explosions of femtosecond laser-heated deuterium clusters},
    journal = {Nature},
    volume = {398},
    year = {1999},
    pages = {489--492},
    DOI = {10.1038/19037}
}

@article{Suzuki18,
  title = {{Time-Dependent Multicomponent Density Functional Theory for Coupled Electron-Positron Dynamics}},
  author = {Suzuki, Yasumitsu and Hagiwara, Satoshi and Watanabe, Kazuyuki},
  journal = {Phys. Rev. Lett.},
  volume = {121},
  pages = {133001},
  year = {2018},
  doi = {10.1103/PhysRevLett.121.133001},
}

@article{Schafer90,
  title={{Energy analysis of time-dependent wave functions: Application to above-threshold ionization}},
  author={Schafer, KJ and Kulander, KC},
  journal={Phys. Rev. A},
  volume={42},
  pages={5794--5797},
  year={1990},
  doi={10.1103/PhysRevA.42.5794}
}

@article{Paulus1994,
  title={Rescattering effects in above-threshold ionization: a classical model},
  author={Paulus, Gerhard G and Becker, Wilhelm and Nicklich, W and Walther, Herbert},
  journal={J. Phys. B},
  volume={27},
  pages={L703},
  year={1994},
  doi={10.1088/0953-4075/27/21/003}
}

@article{Walker94,
  title = {Precision Measurement of Strong Field Double Ionization of Helium},
  author = {Walker, B. and Sheehy, B. and DiMauro, L. F. and Agostini, P. and Schafer, K. J. and Kulander, K. C.},
  journal = {Phys. Rev. Lett.},
  volume = {73},
  issue = {9},
  pages = {1227--1230},
  numpages = {0},
  year = {1994},
  month = {Aug},
  publisher = {American Physical Society},
  doi = {10.1103/PhysRevLett.73.1227},
  url = {https://link.aps.org/doi/10.1103/PhysRevLett.73.1227}
}

@article{Saito2000,
  title={Is positronium hydride atom or molecule?},
  author={Saito, Shiro L},
  journal={Nucl. Instrum. Methods Phys. Res. B},
  volume={171},
  pages={60--66},
  year={2000},
  doi={10.1016/S0168-583X(00)00005-7}
}

@article{saito2003_psh_mrci,
  title={Multireference configuration interaction calculations of some low-lying states of positronium hydride},
  author={Saito, Shiro L},
  journal={J. Chem. Phys.},
  volume={118},
  pages={1714--1720},
  year={2003},
  doi={10.1063/1.1531101}
}

@article{saito2003_mrci_ps_halides,
  title={Calculations for the ground states of positronium halides by singly and doubly excited configuration interaction method},
  author={Saito, Shiro L},
  journal={Chem. Phys. Lett.},
  volume={381},
  pages={565--571},
  year={2003},
  doi={10.1016/j.cplett.2003.10.014}
}

@article{Bromley2001,
  title={{Configuration-interaction calculations of PsH and $e^+$ Be}},
  author={Bromley, Michael W J and Mitroy, Jim},
  journal={Phys. Rev. A},
  volume={65},
  pages={012505},
  year={2001},
  doi={10.1103/PhysRevA.65.012505}
}

@article{Bromley2006,
  title={{Large-dimension configuration-interaction calculations of positron binding to the group-II atoms}},
  author={Bromley, Michael W J and Mitroy, James},
  journal={Phys. Rev. A},
  volume={73},
  number={3},
  pages={032507},
  year={2006},
  doi={10.1103/PhysRevA.73.032507}
}

@article{Becker2018,
  title={The plateau in above-threshold ionization: the keystone of rescattering physics},
  author={Becker, W and Goreslavski, SP and Milo{\v{s}}evi{\'c}, DB and Paulus, G G},
  journal={J. Phys. B},
  volume={51},
  pages={162002},
  year={2018},
  doi={10.1088/1361-6455/aad150}
}

@article{Corkum2007,
    title = {{Attosecond science}},
    volume = {3},
    doi = {10.1038/nphys620},
    journal = {Nat. Phys.},
    author = {Corkum, P. B. and Krausz, F.},
    year = {2007},
    pages = {381--387}
}

@article{Krausz2009,
    title = {{Attosecond physics}},
    volume = {81},
    doi = {10.1103/RevModPhys.81.163},
    journal = {Rev. Mod. Phys.},
    author = {Krausz, F. and Ivanov, M.},
    year = {2009},
    pages = {163--234}
}

@article{Nisoli2017,
    title = {Attosecond {Electron} {Dynamics} in {Molecules}},
    volume = {117},
    doi = {10.1021/acs.chemrev.6b00453},
    journal = {Chem. Rev.},
    author = {Nisoli, M. and Decleva, P. and Calegari, F. and Palacios, A. and Mart{\'{\i}}n, F.},
    year = {2017},
    pages = {10760--10825}
}

@article{Kulander1988,
    title = {{Time-dependent theory of multiphoton ionization of xenon}},
    volume = {38},
    doi = {10.1103/PhysRevA.38.778},
    journal = {Phys. Rev. A},
    author = {Kulander, K. C.},
    year = {1988},
    pages = {778--787}
}

@incollection{Kulander1993,
    address = {Boston, MA},
    series = {{NATO} {ASI} {Series}},
    title = {Dynamics of {Short}-{Pulse} {Excitation}, {Ionization} and {Harmonic} {Conversion}},
    isbn = {978-1-4615-7963-2},
    booktitle = {Super-{Intense} {Laser}-{Atom} {Physics}},
    publisher = {Springer},
    author = {Kulander, K. C. and Schafer, K. J. and Krause, J. L.},
    editor = {Piraux, B. and L’Huillier, A. and Rz{\c a}{\.z}ewski, K.},
    year = {1993},
    doi = {10.1007/978-1-4615-7963-2_10},
    pages = {95--110}
}

@article{Yan1985,
  title={Impulsive stimulated scattering: General importance in femtosecond laser pulse interactions with matter, and spectroscopic applications},
  author={Yan, Yong-Xin and Gamble Jr, Edward B and Nelson, Keith A},
  journal={J. Chem. Phys.},
  volume={83},
  pages={5391--5399},
  year={1985},
  doi={10.1063/1.449708}
}

@article{Barth2009,
    title = {Trigonometric pulse envelopes for laser-induced quantum dynamics},
    volume = {42},
    doi = {10.1088/0953-4075/42/23/235101},
    journal = {J. Phys. B},
    author = {Barth, I and Lasser, C},
    year = {2009},
    pages = {235101}
}

@article{Ehrenfest1927,
    title = {{Bemerkung {\"u}ber die angen{\"a}herte G{\"u}ltigkeit der klassischen Mechanik innerhalb der Quantenmechanik}},
    volume = {45},
    doi = {10.1007/BF01329203},
    journal = {Z. Phys.},
    author = {Ehrenfest, P.},
    year = {1927},
    pages = {455--457}
}

@article{Born1927,
    title = {{Zur Quantentheorie der Molekeln}},
    volume = {389},
    doi = {10.1002/ANDP.19273892002},
    journal = {Ann. Phys.},
    author = {Born, M. and Oppenheimer, R.},
    year = {1927},
    pages = {457--484}
}

@book{Born1954,
    address = {Oxford},
    title = {{Dynamical Theory of Crystal Lattices}},
    publisher = {Clarendon Press},
    author = {Born, Max and Huang, Kun},
    year = {1954}
}

@article{moskal_positronium_2019,
    title = {Positronium in medicine and biology},
    volume = {1},
    doi = {10.1038/s42254-019-0078-7},
    journal = {Nat. Rev. Phys.},
    author = {Moskal, Paweł and Jasi{\' n}ska, Bożena and Stępień, Ewa and Bass, Steven D.},
    year = {2019},
    pages = {527--529}
}

@article{bass_colloquium_2023,
    title = {{Colloquium: Positronium physics and biomedical applications}},
    volume = {95},
    doi = {10.1103/RevModPhys.95.021002},
    journal = {Rev. Mod. Phys.},
    author = {Bass, Steven D. and Mariazzi, Sebastiano and Moskal, Pawel and Stępień, Ewa},
    year = {2023},
    pages = {021002}
}

@article{moskal_positronium_2025,
    title = {{Positronium Imaging: History, Current Status, and Future Perspectives}},
    author = {Moskal, Paweł and Bilewicz, Aleksander and Das, Manish and Huang, Bangyan and Khreptak, Aleksander and Parzych, Szymon and Qi, Jinyi and Rominger, Axel and Seifert, Robert and Sharma, Sushil and Shi, Kuangyu and Steinberger, William M. and Walczak, Rafał and Stępień, Ewa},
    volume = {9},
    doi = {10.1109/TRPMS.2025.3583554},
    journal = {IEEE Trans. Radiat. Plasma Med. Sci.},
    year = {2025},
    pages = {981--1001}
}

@article{jean_positron_1990,
    title = {{Positron annihilation spectroscopy for chemical analysis: A novel probe for microstructural analysis of polymers}},
    volume = {42},
    doi = {10.1016/0026-265X(90)90027-3},
    journal = {Microchem. J.},
    author = {Jean, Y. C.},
    year = {1990},
    pages = {72--102}
}

@article{dlubek_positron_2002,
    title = {{Positron Annihilation Lifetime Spectroscopy (PALS) for Interdiffusion Studies in Disperse Blends of Compatible Polymers: A Quantitative Analysis}},
    volume = {35},
    doi = {10.1021/ma020451z},
    journal = {Macromolecules},
    author = {Dlubek, G. and Pionteck, J. and Bondarenko, V. and Pompe, G. and Taesler, Ch. and Petters, K. and Krause-Rehberg, R.},
    year = {2002},
    pages = {6313--6323}
}

@article{attallah_revisiting_2024,
    title = {{Revisiting Metal–Organic Frameworks Porosimetry by Positron Annihilation: Metal Ion States and Positronium Parameters}},
    volume = {15},
    doi = {10.1021/acs.jpclett.4c00762},
    journal = {J. Phys. Chem. Lett.},
    author = {Attallah, Ahmed G. and Bon, Volodymyr and Maity, Kartik and Zaleski, Radosław and Hirschmann, Eric and Kaskel, Stefan and Wagner, Andreas},
    year = {2024},
    pages = {4560--4567}
}

@article{singh_positron_2016,
    title = {Positron annihilation spectroscopy in tomorrow's material defect studies},
    volume = {51},
    doi = {10.1080/05704928.2016.1141290},
    journal = {Appl. Spectrosc. Rev.},
    author = {Singh, Aditya Narayan},
    year = {2016},
    pages = {359--378}
}

@article{gidley_positron_2006,
    title = {{Positron annihilation as a method to characterize porous materials}},
    volume = {36},
    doi = {10.1146/annurev.matsci.36.111904.135144},
    journal = {Annu. Rev. Mater. Res.},
    author = {Gidley, David W. and Peng, Hua-Gen and Vallery, Richard S.},
    year = {2006},
    pages = {49--79}
}

@article{gribakin_positron-molecule_2010,
    title = {Positron-molecule interactions: {Resonant} attachment, annihilation, and bound states},
    volume = {82},
    doi = {10.1103/RevModPhys.82.2557},
    journal = {Rev. Mod. Phys.},
    author = {Gribakin, G. F. and Young, J. A. and Surko, C. M.},
    year = {2010},
    pages = {2557--2607}
}

@article{mitroy_measuring_1999,
    title = {Measuring the positron affinities of atoms},
    volume = {32},
    doi = {10.1088/0953-4075/32/15/101},
    journal = {J. Phys. B},
    author = {Mitroy, J. and Ryzhikh, G. G.},
    year = {1999},
    pages = {L411}
}

@article{Surko_2012,
doi = {10.1088/1367-2630/14/6/065004},
url = {https://doi.org/10.1088/1367-2630/14/6/065004},
year = {2012},
month = {jun},
publisher = {IOP Publishing},
volume = {14},
number = {6},
pages = {065004},
author = {Surko, C M and Danielson, J R and Gribakin, G F and Continetti, R E},
title = {Measuring positron–atom binding energies through laser-assisted photorecombination},
journal = {New Journal of Physics}
}

\end{document}